\documentclass[fleqn,10pt]{wlscirep}

\usepackage{caption}
\usepackage{comment}
\usepackage[utf8]{inputenc}
\usepackage[T1]{fontenc}
\usepackage[most]{tcolorbox}

\title{A large-scale dataset of Android applications\\and their SDK dependencies}

\author[1,*]{Aurora Gori Savellini}
\author[1,2]{Tiziano Squartini}
\author[1,3]{Massimo Riccaboni}
\affil[1]{IMT School for Advanced Studies, P.zza San Francesco 19, 55100 Lucca (Italy)}
\affil[2]{Istituto Nazionale di Alta Matematica `Francesco Severi' (INdAM-GNAMPA), P.le Aldo Moro 5, 00185 Rome (Italy)}
\affil[3]{IUSS University School for Advanced Studies, P.zza della Vittoria 15, 27100 Pavia (Italy)}
\affil[*]{aurora.gorisavellini@imtlucca.it}

\begin{abstract}
Mobile applications (apps) increasingly rely on third-party Software Development Kits (SDKs) to provide services such as advertising, analytics, authentication, crash reporting, and location services. These components form an important, but often hidden, layer of the mobile ecosystem. Here, we present a large-scale dataset linking Android apps to the third-party SDKs they integrate. The dataset was constructed by combining app package files (APKs) and metadata from AndroZoo with SDK detection rules provided by Exodus Privacy. We implemented a reproducible, static-analysis pipeline that downloads APKs, inspects their compiled code, and detects SDKs through code-level signature matching: the released dataset contains 334,719 unique app-version observations (associated with 99,722 Android apps) and 246 third-party SDKs - including app-level metadata such as categories, comments, file size, download counts, ratings and SDK-level metadata such as categories, code roots, code versions, names. The dataset supports the construction of a large-scale app-SDK dependency network and its projection onto both layers; in addition, SDKs are mapped to their corresponding operating companies, enabling provider-level analysis of technological concentration and upstream control in the mobile ecosystem. To support reproducible research, the pipeline and dataset are publicly released on GitHub and Zenodo; more broadly, the project provides a reusable research infrastructure for studying third-party technological dependencies, data-collection capabilities, and privacy infrastructures across Android apps.
\end{abstract}

\begin{document}

\flushbottom

\maketitle

\thispagestyle{empty}

\section*{Background \& Summary}

Mobile applications (apps) increasingly rely on third-party Software Development Kits (SDKs)\footnote{Software Development Kits (SDKs) are reusable software components distributed by third-party providers that enable app developers to integrate functionalities such as advertising, analytics, user engagement, and other data-driven services without developing them internally. In this study, we focus specifically on tracking SDKs, i.e. SDKs that can support the collection, processing, or sharing of app- and user-generated data across the mobile ecosystem.}: rather than developing these functionalities internally, app developers integrate reusable software modules supplied by upstream technology providers, thus making SDKs a fundamental layer of the Android ecosystem and enabling software reuse, modular production, and the specialization of technological capabilities across firms.

Beyond their engineering role, tracking SDKs also constitute an important data-collection infrastructure. Since they are directly embedded into the app code, they inherit the permissions granted to the host app and can mediate the collection, processing, and exchange of user- and app-generated data: through advertising, analytics, authentication, cloud, and engagement services, SDK providers may aggregate information across thousands of otherwise unrelated applications, thus shaping both software dependencies and data flows throughout the mobile ecosystem.

Despite their growing importance, openly available datasets linking Android apps to their embedded third-party SDKs remain relatively scarce. Existing app-store metadata do not explicitly report SDK dependencies, while commercial mobile intelligence platforms typically provide proprietary analyses under restricted-access models: recovering app-SDK relationships, therefore, requires inspecting Android Package Kit (APK)\footnote{Android Package Kit (APK) files are the standard application package format used to distribute and install Android apps: they contain configuration files, executable code, resources, and embedded third-party software components required for app execution. Throughout the manuscript, the terms `APK', `application package', and `app package file' are used interchangeably.} files and identifying software signatures associated with known SDKs through static analysis.

Here we present a large-scale open dataset linking Android apps distributed through Google Play to the third-party tracking SDKs embedded into their APKs. The dataset was constructed by combining APKs and application metadata retrieved from AndroZoo\footnote{AndroZoo is a large-scale research repository maintained by the University of Luxembourg that provides access to millions of Android application packages (APKs) and associated metadata for scientific research. More information is available at \url{https://androzoo.uni.lu/}} with SDK detection rules provided by Exodus Privacy\footnote{Exodus Privacy is a non-profit organization dedicated to improving transparency about privacy and tracking technologies in mobile applications. It maintains an open database of tracker detection rules and provides tools for static privacy analysis of Android applications. More information is available at \url{https://exodus-privacy.eu.org/en/}.}: the released version contains 334,711 successfully processed APKs corresponding to 99,722 distinct application packages, 246 empirically observed tracking SDKs, 197 provider entities, and more than 845,000 app-SDK relationships. Besides SDK dependencies, the dataset includes rich app-level metadata (e.g. category, download counts, ratings, comments, and file size), together with SDK-level metadata describing software categories, code roots, versions, and provider information.

\begin{figure*}[t]
\centering
\includegraphics[width=\textwidth]{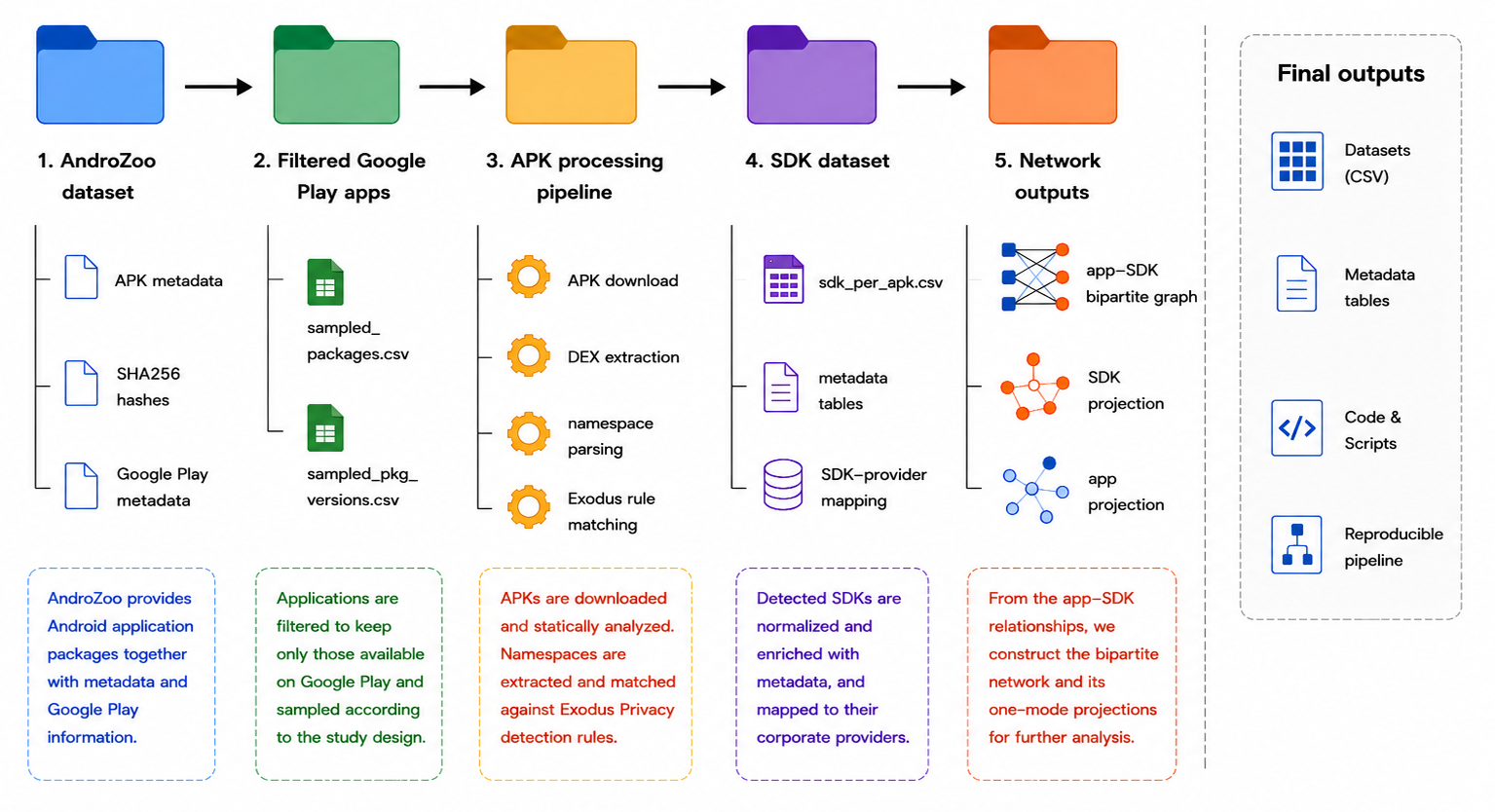}
\caption{{\bf Overview of the dataset construction pipeline.} Starting from the AndroZoo catalog, we retain Android apps observed on Google Play and construct a reproducible sample of APKs and associated versions. APK files are downloaded via the AndroZoo API and processed with static analysis, including DEX extraction, namespace parsing, and matching against Exodus Privacy detection rules. Detected SDKs are normalized, enriched with Google Play metadata, and mapped to their corresponding provider entities. The resulting app-SDK-provider relationships are subsequently represented as bipartite networks and transformed into one-mode projections on the app, SDK, and provider layers. Final released outputs include the app-level dataset, along with the code required to construct network representations and perform downstream analyses.}
\label{fig:dataset_construction_pipeline}
\end{figure*}

The released data naturally support multiple network representations. The primary dataset defines a bipartite app-SDK network in which applications are linked to their embedded SDKs: from this representation, users can construct one-mode projections on both the application and SDK layers, as well as provider-level dependency networks obtained through the mapping of SDKs to their corresponding technology companies; these network representations enable analyses of software dependencies, technological concentration, ecosystem structure, and provider interactions.

This Data Descriptor accompanies both a frozen Zenodo release of the dataset and a fully reproducible GitHub repository. The GitHub repository contains the complete data-construction pipeline, including APK sampling, SDK detection, provider mapping, network construction, community-detection analyses, and technical validation, together with all auxiliary resources required to reproduce the released data - except APK redistribution, which is restricted by the AndroZoo license.

The remainder of this manuscript is organized as follows: the \textit{Methods} section describes the complete data-construction pipeline, including APK acquisition, SDK detection, provider mapping, network construction, and validation procedures; the \textit{Data Records} section documents the released datasets, metadata tables, and network files; the \textit{Technical Validation} section evaluates the completeness, consistency, and structural properties of the released data; the \textit{Data Availability} and \textit{Code Availability} sections describe access to the archived datasets and source code.

\section*{Methods}

\subsection*{Overview of the data construction pipeline}

Figure~\ref{fig:dataset_construction_pipeline} summarizes the workflow used to construct the dataset. The pipeline consists of six main stages: \textit{(i)} acquisition of APKs and metadata from the AndroZoo repository; \textit{(ii)} filtering and sampling Google Play apps\footnote{Although this choice excludes applications distributed exclusively through alternative Android marketplaces, it does not constitute a major practical limitation because Google Play is, by far, the largest official distribution platform for Android apps, therefore providing a broad coverage of the Android ecosystem.}, since AndroZoo provides standardized app metadata for Google Play packages, enabling the integration of app-level characteristics such as category, downloads, ratings, and other descriptive variables; \textit{(iii)} APK processing through static analysis, including APK retrieval, DEX extraction, namespace parsing, and SDK detection based on Exodus Privacy rules; \textit{(iv)} construction of the SDK dataset through SDK normalization, provider mapping, and metadata integration; \textit{(v)} generation of app-SDK and app-provider network representations together with their corresponding one-mode projections; and \textit{(vi)} technical validation procedures aimed at assessing dataset quality, provider attribution, SDK detection coverage, and the robustness of the resulting network representations.

\begin{table*}[t!]
\centering
\renewcommand{\arraystretch}{1.50}
\resizebox{\textwidth}{!}{%
\begin{tabular}{l l r r r r l}
\specialrule{1.2pt}{0pt}{0pt}
\textbf{sha256} & \textbf{pkg\_name} & \textbf{vercode} & \textbf{dex\_date} & \textbf{apk\_size} & \textbf{vt} & \textbf{markets} \\
\specialrule{0.8pt}{0pt}{0pt}
\texttt{0000003B455A...} & \texttt{com.zte.bamachaye} & 121 & 2016-04-05 & 10386469 & 0 & \texttt{anzhi} \\
\specialrule{0.3pt}{0pt}{0pt}
\texttt{000000568FE4...} & \texttt{com.calistree.calistree} & 357 & 1981-01-01 & 43545076 & 0 &
\texttt{play.google.com} \\
\specialrule{0.3pt}{0pt}{0pt}
\texttt{0000014A634D...} & \texttt{com.tanersenel.onlinetvizle} & 16 & 2014-08-20 & 3537486 & 0 & \texttt{PlayDrone} \\
\specialrule{0.3pt}{0pt}{0pt}
\texttt{000001A94F46...} & \texttt{com.firstchoice.myfirstchoice} & 1206145 & 1980-01-01 & 52469861 & 0 & \texttt{play.google.com} \\
\specialrule{0.3pt}{0pt}{0pt}
\texttt{000002B63FAD...} & \texttt{com.deperu.sitiosarequipa} & 10000 & 1980-01-01 & 4300370 & 0 & \texttt{play.google.com} \\
\specialrule{1.2pt}{0pt}{0pt}
\end{tabular}}
\caption{{\bf Illustrative records from the AndroZoo metadata catalog used in this work.} \texttt{sha256} is the unique SHA-256 hash identifying each APK file; \texttt{pkg\_name} denotes the Android application package identifier; \texttt{vercode} is the Android version code associated with the APK release; \texttt{dex\_date} reports the DEX compilation timestamp extracted from the APK metadata; \texttt{apk\_size} indicates the APK file size (in bytes); \texttt{vt} denotes the number of VirusTotal detection engines flagging the APK as potentially malicious; and \texttt{markets} lists the application marketplace(s) from which the APK was collected by AndroZoo. Placeholder timestamps such as \texttt{1980-01-01} and \texttt{1981-01-01} are known artifacts of the AndroZoo metadata infrastructure and do not correspond to actual APK release dates.}
\label{tab:androzoo_example_rows}
\end{table*}

\subsection*{Source data: AndroZoo metadata catalog and APKs}

The AndroZoo metadata catalog is distributed as a compressed comma-separated values (CSV) archive and updated continuously. The catalog was downloaded on October 7, 2025, the \texttt{latest.csv.gz} file contained 26,301,149 APK observations, of which 23,348,168 (approximately 88.8\%) were associated with Google Play: each row corresponds to a specific APK observation and includes core metadata fields such as \texttt{sha256}, \texttt{sha1}, \texttt{md5}, \texttt{pkg\_name}, \texttt{vercode}, \texttt{dex\_date}, \texttt{dex\_size}, \texttt{apk\_size}, \texttt{vt\_detection}, \texttt{vt\_scan\_date}, and \texttt{markets}. Table~\ref{tab:androzoo_example_rows} reports illustrative rows from the catalog. Let us, now, provide a detailed description of each field:

\begin{itemize}
\item \texttt{sha256}, \texttt{sha1}, and \texttt{md5} provide cryptographic identifiers for each APK;
\item \texttt{pkg\_name} and \texttt{vercode} identify the Android package and version code reported in the application manifest;
\item \texttt{dex\_size}\footnote{AndroZoo documentation notes that the \texttt{dex\_date} field may be occasionally unreliable, particularly for Google Play applications. Placeholder timestamps such as \texttt{1980-01-01} and \texttt{1981-01-01} are frequently observed in the metadata and do not correspond to actual Android release dates: as a consequence, \texttt{dex\_date} should be interpreted as an approximate temporal indicator rather than an exact release timestamp.} and \texttt{apk\_size} report the size of the main DEX file and of the APK, respectively;
\item \texttt{vt\_detection} and \texttt{vt\_scan\_date} report the number of VirusTotal engines flagging the APK as potentially malicious and the corresponding scan date;
\item \texttt{markets} reports the application markets in which AndroZoo observed the APK. 
\end{itemize}

In the present work, the AndroZoo catalog served both as a metadata source and as the basis for large-scale APK retrieval: access to APK files was performed through the official AndroZoo API, by using authenticated API requests and SHA256-based APK identifiers. APK files were downloaded dynamically and processed entirely in temporary memory buffers without permanent storage of the raw application binaries - a design that has substantially reduced storage requirements while enabling scalable static analysis of hundreds of thousands of APK files. Parallel downloads and APK processing were executed using a controlled multi-threaded architecture compliant with AndroZoo usage recommendations regarding concurrent requests.

The AndroZoo catalog combines information concerning specific APK versions with metadata collected at the app level: some metadata fields may, therefore, reflect the state of an application at the time the metadata were acquired rather than at the original APK release date; as a consequence, variables such as descriptions, download counts, or ratings should be interpreted with reference to the metadata acquisition date, when available.

\begin{figure}[t!]
\centering
\includegraphics[width=0.75\linewidth]{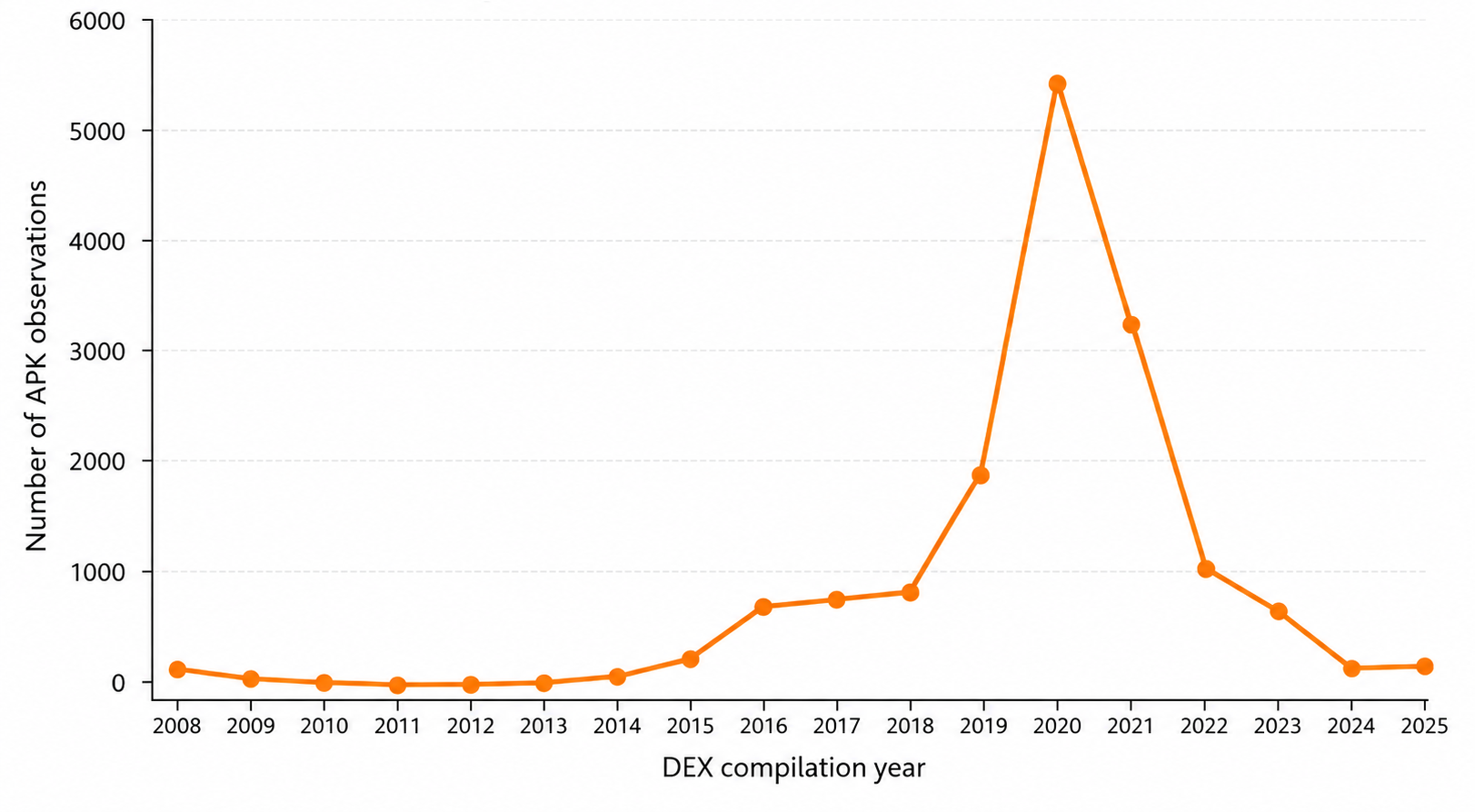}
\caption{{\bf Temporal distribution of APK observations.} The yearly distribution of APK observations reflects the temporal composition of the underlying AndroZoo repository, from which APKs were randomly sampled: as a consequence, the peak around 2020 should be interpreted as a characteristic of the available APK on Androzoo rather than as a direct indication of Android ecosystem dynamics.}
\label{fig:temporal_distribution_apk}
\end{figure}

\subsection*{Sampling Google Play apps}

The construction pipeline first filtered the AndroZoo catalog to retain only apps observed on Google Play~\cite{Allix:2016:ACM:2901739.2903508}: specifically, apps were selected when the \texttt{markets} metadata field contained the string \texttt{play.google.com}—a restriction introduced to improve comparability across apps by focusing on a relatively homogeneous distribution environment.

After filtering the AndroZoo catalog, APKs were randomly sampled, without stratification by app category, download counts, or other metadata attributes. Random sampling was adopted to avoid introducing selection biases and to preserve, as closely as possible, the statistical properties of the underlying population of Google Play APKs available in AndroZoo. A fixed random seed was used to ensure full reproducibility of the selection procedure.

As shown in Figure~\ref{fig:temporal_distribution_apk}, the temporal distribution of the sampled APKs is concentrated in recent years. Since the released dataset was obtained through random sampling from the Google Play APKs available in AndroZoo, it naturally inherits the temporal composition of the underlying repository rather than imposing an artificial temporal balance. Consequently, the peak around 2020 should be interpreted as a characteristic of the available APK corpus in AndroZoo rather than as direct evidence of a specific temporal trend in the Android ecosystem\footnote{Researchers interested in temporally balanced analyses may construct year-specific subsamples or temporally stratified network representations using the released \texttt{dex\_date} metadata.}.

The pipeline was designed to target approximately 100,000 Android apps while preserving all available APKs versions associated with each sampled package: consequently, the final dataset contains multiple historical versions for many apps, enabling both cross-sectional and longitudinal analyses of SDK adoption and dependency evolution over time. To identify valid applications before APK downloading, the pipeline queried the AndroZoo Google Play metadata API and retrieved version-code information associated with each package: applications without retrievable metadata or valid APK-version mappings were excluded from the final sample.

The \texttt{dex\_date} metadata field provided by AndroZoo was inspected to obtain an approximate temporal characterization of the dataset: after excluding dates outside the historical period of the Android ecosystem, the remaining valid timestamps span the period from 2008-02-28 to 2025-10-05.

\begin{figure*}[t!]
\centering
\includegraphics[width=\textwidth]{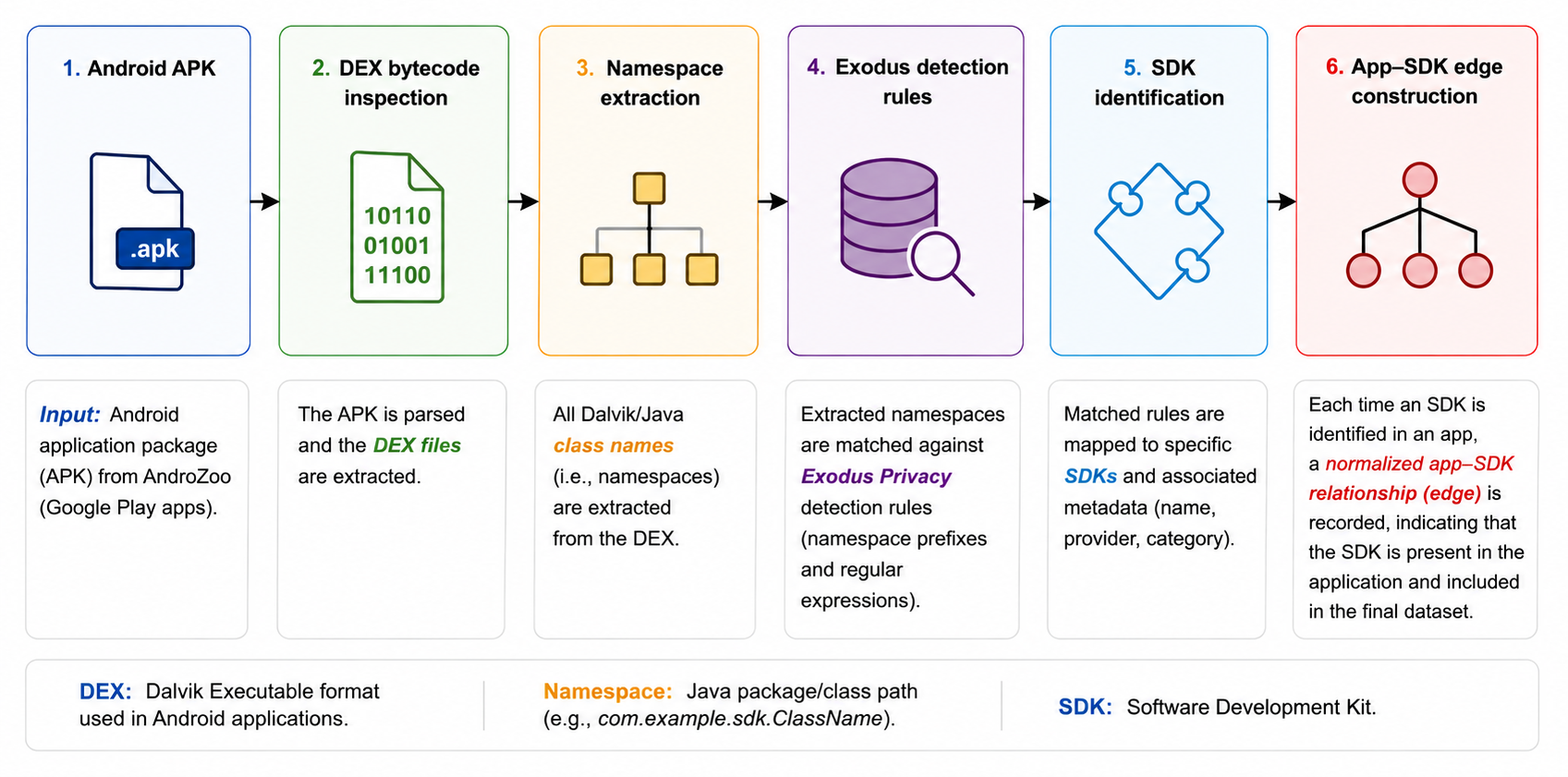}
\caption{{\bf Overview of the tracker SDK detection workflow used to construct the app-SDK dataset.} Android APKs are processed through static analysis of DEX bytecode, from which Dalvik/Java namespaces are extracted. Extracted namespaces are matched against publicly available Exodus Privacy detection rules based on namespace prefixes and regular-expression patterns. Matched signatures are mapped to SDK identifiers and associated metadata, including provider names and functional categories. The final output is a normalized app-SDK edge representation suitable for constructing bipartite dependency networks and derived network projections.}
\label{fig:sdk_detection_workflow}
\end{figure*}

\subsection*{SDK detection rules from Exodus Privacy}

SDK detection was performed using publicly available tracking and SDK signature rules provided by the Exodus Privacy project, an open-source initiative aimed at identifying third-party trackers embedded into Android apps. Exodus Privacy maintains a publicly documented repository of Android tracking SDK signatures together with associated tracker descriptions, functional categories, and code-level detection rules.

Each tracker entry includes metadata describing the SDK provider, the functional role of the tracker (advertising, analytics, authentication, crash reporting, location services, etc.), and one or more namespace-based detection signatures used for static analysis of Android APKs. The Exodus rules consist primarily of namespace-prefix signatures and regular-expression patterns associated with known SDK providers and tracking libraries. These rules were parsed and normalized into two complementary representations, namely prefix-based namespace rules and regular-expression matching rules operating on Dalvik class paths.

The SDK identification workflow consists of six main steps: \textit{(i)} retrieval of Android APK files from the AndroZoo repository; \textit{(ii)} parsing of each APK archive and extraction of the compiled DEX bytecode; \textit{(iii)} extraction of all Dalvik/Java class namespaces from the DEX files; \textit{(iv)} matching of extracted namespaces against the Exodus Privacy detection rules using both namespace-prefix and regular-expression signatures; \textit{(v)} mapping of matched signatures to normalized SDK identifiers together with the associated metadata (e.g. provider name and functional category); and \textit{(vi)} construction of a normalized app--SDK edge list, which constitutes the basis for the released bipartite network and its corresponding one-mode projections.

The Exodus detection methodology relies on static inspection of APK bytecode rather than app decompilation or runtime execution. Namespace matching was performed on normalized lowercase Dalvik class paths: an SDK was considered detected when at least one extracted namespace matched either a namespace-prefix rule or a regular-expression signature associated with the SDK. Namespace roots were subsequently normalized through heuristic consolidation rules designed to reduce fragmentation caused by nested internal package structures; no de-obfuscation or runtime unpacking procedures were applied. The resulting rule base included 588 namespace-prefix rules and 432 regular-expression rules; whenever available, each rule was associated with SDK names and functional categories.

Since these SDKs are embedded into the app code, thereby inheriting the permissions granted to the host application, their integration may enable the collection, mediation, or processing of user- and app-generated information by upstream SDK providers. The resulting dataset therefore captures large-scale third-party software dependency structures associated with tracking and analytics infrastructures embedded into Android apps. Using publicly documented SDK signatures allows consistent identification of SDK dependencies across large collections of APKs.

As with most static-analysis approaches, heavily obfuscated, dynamically loaded, or aggressively renamed SDK code may reduce detection accuracy and contribute to false negatives; conversely, because detection relies on static namespace matching, the presence of SDK code within an APK does not necessarily imply active runtime execution or realized data transmission during app use.

\subsection*{SDK extraction and normalization}

For each successfully processed APK, the pipeline generated a normalized set of detected SDKs together with associated namespace roots and functional categories: since SDKs may appear through multiple nested namespaces or internal package structures, extracted namespaces were normalized to canonical SDK roots whenever possible; additional heuristics were applied to reduce namespace fragmentation and eliminate generic package prefixes that, otherwise, could artificially inflate SDK counts.

The pipeline distinguishes between raw namespace matches extracted from bytecode, normalized SDK identifiers, canonical provider identifiers, and SDK functional categories. Table~\ref{tab:sdk_dataset_example} reports illustrative rows from the final dataset and provides a concrete example of how detected SDKs and associated metadata are represented for each processed APK.

The final dataset contains 246 empirically observed normalized SDK entities connected through 845,010 app-SDK edges.

\subsection*{SDK-provider mapping}

To enable provider-level analyses, detected SDKs were mapped to their corresponding operating companies using a hybrid automated and manually validated procedure. The starting point was the list of SDKs and tracker pages provided by Exodus Privacy: for each SDK, we extracted the tracker name, functional category, Exodus URL, and, when available, the official tracker website linked from the Exodus report page. Table~\ref{tab:provider_mapping_variables} summarizes the main variables included in the final canonical SDK-provider mapping dataset.

Provider attribution was first performed automatically. The pipeline queried the official tracker website and extracted candidate company names from multiple public sources, including website footers, metadata fields, structured JSON-LD data, legal or privacy pages, and textual information reported on Exodus tracker pages. Candidate legal names were identified using regular-expression patterns for common corporate suffixes such as \texttt{Inc.}, \texttt{LLC}, \texttt{Ltd.}, \texttt{GmbH}, \texttt{S.p.A.}, and related forms. This automated step recovered company information for 57.41\% of SDK entries.

\begin{table*}[t!]
\centering
\renewcommand{\arraystretch}{1.50}
\resizebox{\textwidth}{!}{%
\begin{tabular}{l l r r l}
\specialrule{1.2pt}{0pt}{0pt}
\textbf{sha256} & \textbf{pkg\_name} & \textbf{vercode} & \textbf{sdk\_count} & \textbf{sdk\_names} \\
\specialrule{0.8pt}{0pt}{0pt}
\texttt{0002B476CAF9...} & \texttt{com.rssinteractive.paratic.haber} & 16 & 2 & \texttt{Google Crashlytics; Google Firebase Analytics} \\
\specialrule{0.3pt}{0pt}{0pt}
\texttt{0002F1B36468...} & \texttt{com.jui.snakewallpaper} & 2 & 2 & \texttt{Google AdMob; Google Firebase Analytics} \\
\specialrule{0.3pt}{0pt}{0pt}
\texttt{00001B70E724...} & \texttt{com.fr.quizzapp} & 7 & 4 & \texttt{Google AdMob; Google Firebase Analytics; Facebook Login; Unity Ads} \\
\specialrule{0.3pt}{0pt}{0pt}
\texttt{00010A2579FE...} & \texttt{com.stylish.font.neonkeyboard} & 14 & 2 & \texttt{Google AdMob; Google Firebase Analytics} \\
\specialrule{0.3pt}{0pt}{0pt}
\texttt{0000C0045C36...} & \texttt{com.egyking.ancientgreece} & 3 & 2 &
\texttt{Google AdMob; Google Firebase Analytics} \\
\specialrule{1.2pt}{0pt}{0pt}
\end{tabular}}
\caption{{\bf Illustrative rows from the final app-SDK dataset.} Let us explicitly notice that each row corresponds to one processed APK and reports the detected SDKs associated with that app version. Additional variables released with the dataset include processing status indicators, extraction timestamps, normalized namespace roots, provider identifiers, and functional categories.}
\label{tab:sdk_dataset_example}
\end{table*}

For SDKs without reliable automated attribution, provider information was completed through a curated manual mapping procedure. Manual assignments followed a standardized protocol and were based on multiple independent sources, including official SDK documentation, developer documentation, corporate websites, privacy policies, acquisition announcements, and public records describing ownership relationships between SDK products and firms. Whenever multiple sources were available, provider ownership was assigned only when consistent evidence supported the attribution, while official corporate documentation and acquisition records were treated as authoritative in cases of conflicting information. This step accounted for 42.59\% of SDK entries not reliably resolved through automated extraction.

\begin{table*}[t!]
\centering
\renewcommand{\arraystretch}{1.50}
\resizebox{\textwidth}{!}{%
\begin{tabular}{l l}
\specialrule{1.2pt}{0pt}{0pt}
\textbf{Variable} & \textbf{Description} \\
\specialrule{0.8pt}{0pt}{0pt}
\texttt{sdk\_name} & Original SDK name from the Exodus Privacy tracker database. \\
\specialrule{0.3pt}{0pt}{0pt}
\texttt{sdk\_type} & Functional category assigned by Exodus Privacy (e.g., advertising, analytics, location). \\
\specialrule{0.3pt}{0pt}{0pt}
\texttt{tracker\_website} & Official website of the SDK or vendor. \\
\specialrule{0.3pt}{0pt}{0pt}
\texttt{declared\_company} & Company name automatically extracted from public sources. \\
\specialrule{0.3pt}{0pt}{0pt}
\texttt{provider\_manual\_assignment} & Manual provider assignment used when automated extraction was missing or inaccurate. \\
\specialrule{0.3pt}{0pt}{0pt}
\texttt{provider\_validated} & Validated provider name after manual correction of noisy scraped entries. \\
\specialrule{0.3pt}{0pt}{0pt}
\texttt{provider\_raw} & Reconciled provider name obtained by combining all available sources. \\
\specialrule{0.3pt}{0pt}{0pt}
\texttt{provider\_source} & Source ultimately used for provider attribution. \\
\specialrule{0.3pt}{0pt}{0pt}
\texttt{company\_canonical} & Final standardized provider label used in provider-level analyses. \\
\specialrule{1.2pt}{0pt}{0pt}
\end{tabular}}
\caption{{\bf Core variables included in the final canonical SDK-provider mapping dataset.} Let us explicitly notice that the variable \texttt{company\_canonical} harmonizes aliases, acquired firms, and subsidiaries under a common provider label: for example, \texttt{Google AdMob}, \texttt{Google Firebase Analytics}, and \texttt{Google Tag Manager} are all mapped to \texttt{Google}.}
\label{tab:provider_mapping_variables}
\end{table*}

Manual validation was primarily required for SDKs associated with corporate acquisitions, mergers, rebranding events, discontinued products, legacy libraries, or ambiguous product-level naming conventions. When SDK ownership could be clearly attributed to a parent company through acquisitions or corporate integration, the canonical provider label was assigned to the acquiring firm (e.g., Firebase $\rightarrow$ Google; MoPub $\rightarrow$ AppLovin). Provider ownership attribution reflects the corporate ownership structure observable at the time the mapping procedure was conducted.

Finally, company names were canonicalized to reduce duplication arising from aliases, subsidiaries, spelling variants, legal suffixes, rebranding events, and product-level labels. For example, Firebase-related entries and Google developer services were mapped to a common Google provider label, while Facebook- and Meta-related entries were mapped to a common Meta provider label. Candidate duplicates were initially identified through normalized company names and fuzzy string matching and subsequently verified manually before consolidation. The complete SDK-to-provider mapping is released together with the dataset, allowing users to inspect, verify, and extend provider assignments as ownership relationships evolve over time.

The final provider mapping achieves complete coverage of all SDK entries extracted from the Exodus Privacy tracker database, linking 431 raw SDK labels and aliases to 338 canonical provider identities. After SDK normalization and filtering, the final detectable rule base used for APK analysis contains 277 distinct tracker signatures, of which 246 are observed within the released APK sample\footnote{The remaining detectable tracker signatures correspond to low-prevalence, obsolete, potentially obfuscated, region-specific, or sparsely distributed libraries that do not appear within the filtered sample of successfully processed Google Play apps.}.

\begin{figure*}[t!]
\centering
\includegraphics[width=\textwidth]{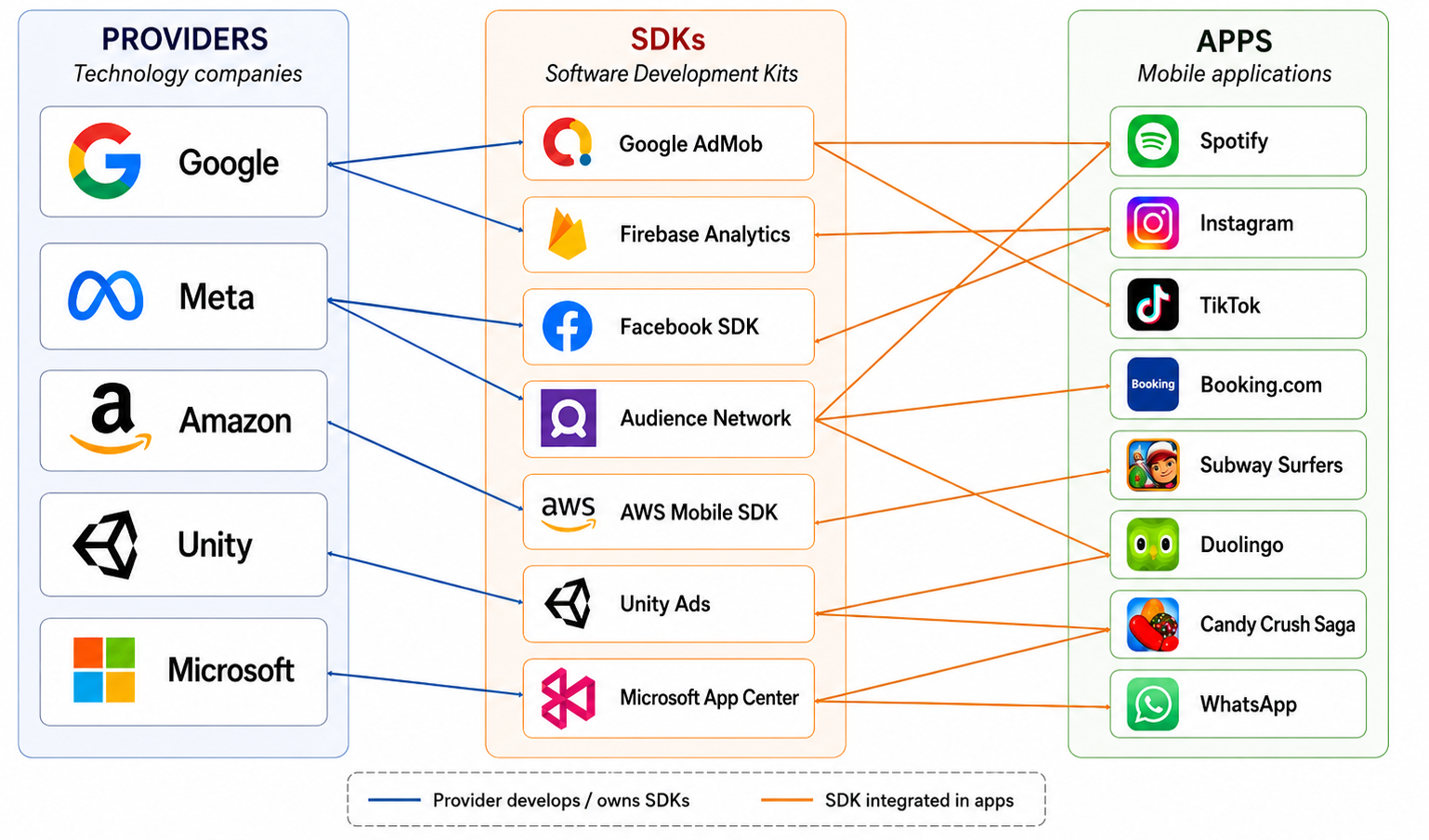}
\caption{{\bf Illustrative tripartite provider-SDK-app dependency structure.} Technology providers (left) develop and maintain SDKs (center), which are embedded into mobile apps (right). The app-SDK and app-provider networks released in this dataset can be interpreted as subsets of this underlying tripartite structure. The entities shown are illustrative examples and do not represent the complete dataset.}
\label{fig:tripartite_structure}
\end{figure*}

This mapping reveals that the mobile tracking ecosystem combines a long tail of specialized firms with a relatively small group of providers controlling multiple SDKs. While most companies are associated with a single software component, several large technology and advertising firms maintain extensive portfolios of SDKs. Table~\ref{tab:top_providers} summarizes the providers associated with the largest number of distinct SDKs in the released dataset. This concentration highlights the role of SDK ecosystems as modular infrastructures through which a limited number of dominant firms extend their tracking and analytics infrastructures across a large population of third-party mobile applications. Figure~\ref{fig:tripartite_structure} provides an illustrative representation of the tripartite provider--SDK--app dependency structure underlying the released dataset.

\subsection*{Construction of the app-SDK and the app-provider networks}

The final app-SDK dataset was constructed by combining app-level identifiers and metadata obtained from AndroZoo with SDK information extracted through static analysis of APKs: SDK dependencies can be summed up into an app-SDK edge list, each row of which indicates the presence of a specific SDK within a given app. 

Such an edge list naturally defines a bipartite app-SDK network in which apps are connected to the SDKs embedded into their code: more formally, we can indicate such a network as $\mathbf{G}=\{\mathcal{A},\mathcal{S},\mathcal{E}\}$, where $\mathcal{A}$ denotes the set of apps, $\mathcal{S}$ denotes the set of detected SDKs, and $\mathcal{E}$ denotes the set of empirical app-SDK dependencies, or, in the jargon of network science, edges: more explicitly, the edge $(i,\alpha)$ is present, i.e. $(i,\alpha)\in\mathcal{E}$, if app $i$ embeds SDK $\alpha$ within the corresponding APK.

Upon letting $N=|\mathcal{A}|$ denote the number of apps and $M=|\mathcal{S}|$ denote the number of SDKs, the bipartite network above remains unambiguously identified by the $N\times M$ biadjacency matrix $\mathbf{B}=\{b_{i\alpha}\}$, the generic entry of which is defined as

\begin{equation}
b_{i\alpha}=
\begin{cases}
1, & \text{if application\:} i \text{\:embeds SDK\:} \alpha,\\
0, & \text{otherwise.}
\end{cases}
\end{equation}

When multiple APK versions are available for the same package, the released dataset preserves these observations rather than collapsing them into a single record, a design allowing the temporal evolution of SDK dependencies to be retained and enabling network construction at different levels of aggregation: at the most disaggregated level, each observation corresponds to a specific app-version pair, allowing the construction of a bipartite network in which distinct versions of the same app are treated as separate nodes; alternatively, aggregation can be carried out at the package level, by combining all versions associated with the same app into a single node. Here, we have opted for aggregating all versions of the same app into a single node.

The degree of an app node is defined as

\begin{equation}
k_i=\sum_{\alpha=1}^Mb_{i\alpha},
\end{equation}
corresponding to the number of SDKs embedded into app $i$: this quantity captures the technological complexity of an app's third-party dependency structure and may be interpreted as a proxy for the extent of third-party tracking and analytics integration. Similarly, the degree of an SDK node is defined as

\begin{equation}
h_\alpha=\sum_{i=1}^Nb_{i\alpha},
\end{equation}
corresponding to the number of apps embedding SDK $\alpha$: this quantity measures SDK prevalence within the observed ecosystem and may be interpreted as a proxy for ecosystem penetration or technological reach. Naturally, the total number of app-SDK dependencies is given by $|\mathcal{E}|=L=\sum_{i=1}^Nk_i=\sum_{\alpha=1}^Mh_\alpha$ and the edge density of the ecosystem, usable to compare filtered or temporally restricted versions of the dataset, is given by

\begin{equation}
c=\frac{L}{NM}.
\end{equation}

Starting from $\mathbf{B}$, two network projections remain naturally defined. The app-app projection is defined as the (network represented by the) $N\times N$ adjacency matrix 

\begin{equation}
\mathbf{U}=\mathbf{B}\mathbf{B}^T,
\end{equation}
with $u_{ij}$ denoting the number of SDKs shared by apps $i$ and $j$: the generic edge of the corresponding unweighted representation is 1 if and only if $u_{ij}>0$, thus recording whether the two apps share at least one SDK; analogously, the SDK-SDK projection is defined as the (network represented by the) $M\times M$ adjacency matrix 

\begin{equation}
\mathbf{V}=\mathbf{B}^T\mathbf{B},
\end{equation}
with $v_{\alpha\beta}$ denoting the number of apps in which SDKs $\alpha$ and $\beta$ co-occur: the generic edge of the corresponding unweighted representation is 1 if and only if $v_{\alpha\beta}>0$, thus recording whether the two SDKs co-occur in at least one application.

\begin{table}[t!]
\centering
\caption{\textbf{Providers associated with the largest number of distinct tracking SDKs in the released dataset.}}
\label{tab:top_providers}
\begin{tabular}{lc}
\toprule
\textbf{Provider} & \textbf{Distinct SDKs} \\
\midrule
Google & 10 \\
Meta & 9 \\
Baidu & 8 \\
Tencent & 7 \\
Unity & 6 \\
Amazon & 5 \\
Digital Turbine & 5 \\
Adobe & 5 \\
Alibaba & 4 \\
AppLovin & 4 \\
Microsoft & 4 \\
\bottomrule
\end{tabular}
\end{table}

Since the released dataset includes both dependency information and app-level metadata, researchers can define a wide range of weighting schemes based on attributes such as app size, download counts, review counts, user ratings, and other observable characteristics: for example, edges in projected networks may be weighted according to the cumulative number of downloads associated with shared SDKs or by other measures intended to approximate the scale, popularity, or economic and privacy relevance of the connected applications; the most appropriate weighting strategy depends on the substantive research question and the specific mechanism under investigation.

Let us finally notice that a similar analysis can be carried out using the released SDK-provider mapping, unambiguously identified by the $N\times P$ biadjacency matrix $\mathbf{D}=\{d_{i\alpha}\}$, the generic entry of which is defined as

\begin{equation}
d_{i\alpha} =
\begin{cases}
1, & \text{if application\:} i \text{\:embeds at least one SDK controlled by provider\:} \alpha,\\
0, & \text{otherwise}
\end{cases}
\end{equation}
and leading to the provider-provider projection defined as the (network represented by the) $P\times P$ adjacency matrix \begin{equation}
\mathbf{W}=\mathbf{D}^T\mathbf{D},
\end{equation}
with $w_{\alpha\beta}$ denoting the number of apps in which SDKs controlled by providers $\alpha$ and $\beta$ co-occur.

\begin{table*}[t!]
\centering
\renewcommand{\arraystretch}{1.35}
\resizebox{\textwidth}{!}{%
\begin{tabular}{l l}
\specialrule{1.2pt}{0pt}{0pt}
\textbf{Variable} & \textbf{Description} \\
\specialrule{0.8pt}{0pt}{0pt}
\texttt{sha256} & SHA256 hash uniquely identifying the APK file in AndroZoo. \\
\specialrule{0.3pt}{0pt}{0pt}
\texttt{pkg\_name} & Android package name identifying the application. \\
\specialrule{0.3pt}{0pt}{0pt}
\texttt{vercode} & APK version code from the AndroZoo catalog. \\
\specialrule{0.3pt}{0pt}{0pt}
\texttt{dex\_date} & DEX timestamp associated with the APK; may include placeholder dates. \\
\specialrule{0.3pt}{0pt}{0pt}
\texttt{markets} & Markets in which the APK was observed by AndroZoo. \\
\specialrule{0.3pt}{0pt}{0pt}
\texttt{vt\_detection} & Number of VirusTotal engines flagging the APK at scan time. \\
\specialrule{0.3pt}{0pt}{0pt}
\texttt{vt\_scan\_date} & VirusTotal scan date reported in the AndroZoo catalog. \\
\specialrule{0.3pt}{0pt}{0pt}
\texttt{download\_status} & Status of APK download and static analysis. \\
\specialrule{0.3pt}{0pt}{0pt}
\texttt{extraction\_ts\_utc} & UTC timestamp of SDK extraction. \\
\specialrule{0.3pt}{0pt}{0pt}
\texttt{sdk\_count} & Number of SDKs detected in the APK. \\
\specialrule{0.3pt}{0pt}{0pt}
\texttt{sdk\_names\_json} & JSON list of detected SDK names. \\
\specialrule{0.3pt}{0pt}{0pt}
\texttt{sdk\_roots\_json} & JSON list of namespace roots associated with detected SDKs. \\
\specialrule{0.3pt}{0pt}{0pt}
\texttt{sdk\_types\_json} & JSON list of SDK functional categories. \\
\specialrule{0.3pt}{0pt}{0pt}
\texttt{gp\_title} & Application title retrieved during APK processing. \\
\specialrule{0.3pt}{0pt}{0pt}
\texttt{gp\_source} & Source used for Google Play metadata retrieval. \\
\specialrule{0.3pt}{0pt}{0pt}
\texttt{match\_type} & Type of match between APK version and Google Play metadata. \\
\specialrule{0.3pt}{0pt}{0pt}
\texttt{dex\_date\_meta} & DEX timestamp carried into the metadata table. \\
\specialrule{0.3pt}{0pt}{0pt}
\texttt{gp\_source\_meta} & Google Play metadata source after dataset merging. \\
\specialrule{0.3pt}{0pt}{0pt}
\texttt{match\_type\_meta} & Metadata match type after dataset merging. \\
\specialrule{0.3pt}{0pt}{0pt}
\texttt{meta\_title} & Application title from Google Play metadata. \\
\specialrule{0.3pt}{0pt}{0pt}
\texttt{meta\_descriptionShort} & Short application description from Google Play metadata. \\
\specialrule{0.3pt}{0pt}{0pt}
\texttt{meta\_descriptionHtml} & Full HTML application description. \\
\specialrule{0.3pt}{0pt}{0pt}
\texttt{meta\_az\_metadata\_date} & Date on which AndroZoo acquired the Google Play metadata. \\
\specialrule{0.3pt}{0pt}{0pt}
\texttt{meta\_details.appDetails.packageName} & Package name reported in Google Play metadata. \\
\specialrule{0.3pt}{0pt}{0pt}
\texttt{meta\_details.appDetails.versionCode} & Version code reported in Google Play metadata. \\
\specialrule{0.3pt}{0pt}{0pt}
\texttt{meta\_details.appDetails.versionString} & Human-readable application version string. \\
\specialrule{0.3pt}{0pt}{0pt}
\texttt{meta\_details.appDetails.numDownloads} & Download count category reported in Google Play metadata. \\
\specialrule{0.3pt}{0pt}{0pt}
\texttt{meta\_aggregateRating.starRating} & Average user rating. \\
\specialrule{0.3pt}{0pt}{0pt}
\texttt{meta\_aggregateRating.ratingsCount} & Total number of ratings. \\
\specialrule{0.3pt}{0pt}{0pt}
\texttt{meta\_aggregateRating.commentCount} & Number of comments or reviews. \\
\specialrule{0.3pt}{0pt}{0pt}
\texttt{meta\_details.appDetails.installationSize} & Application installation size. \\
\specialrule{0.3pt}{0pt}{0pt}
\texttt{meta\_details.appDetails.containsAds} & Indicator for whether the app is marked as containing ads. \\
\specialrule{0.3pt}{0pt}{0pt}
\texttt{meta\_details.appDetails.developerName} & Developer name reported in Google Play metadata. \\
\specialrule{0.3pt}{0pt}{0pt}
\texttt{meta\_details.appDetails.developerEmail} & Developer contact email. \\
\specialrule{0.3pt}{0pt}{0pt}
\texttt{meta\_details.appDetails.developerWebsite} & Developer website. \\
\specialrule{0.3pt}{0pt}{0pt}
\texttt{meta\_details.appDetails.developerAddress} & Developer address, when available. \\
\specialrule{0.3pt}{0pt}{0pt}
\texttt{meta\_details.appDetails.uploadDate} & Upload date reported for the application version. \\
\specialrule{0.3pt}{0pt}{0pt}
\texttt{meta\_details.appDetails.recentChangesHtml} & HTML changelog or recent changes field. \\
\specialrule{0.3pt}{0pt}{0pt}
\texttt{meta\_relatedLinks.privacyPolicyUrl} & Application privacy policy URL. \\
\specialrule{0.3pt}{0pt}{0pt}
\texttt{meta\_shareUrl} & Google Play sharing URL for the application. \\
\specialrule{0.3pt}{0pt}{0pt}
\texttt{meta\_details.appDetails.permission.*} & Flattened Android permission fields. \\
\specialrule{0.3pt}{0pt}{0pt}
\texttt{metadata\_merge\_status} & Final indicator reporting whether metadata were matched to the APK observation. \\
\specialrule{1.2pt}{0pt}{0pt}
\end{tabular}}
\caption{Main app-level and SDK-level variables included in the released dataset.}
\label{tab:metadata_variables}
\end{table*}

\begin{figure*}[t!]
\centering
\includegraphics[width=\textwidth]{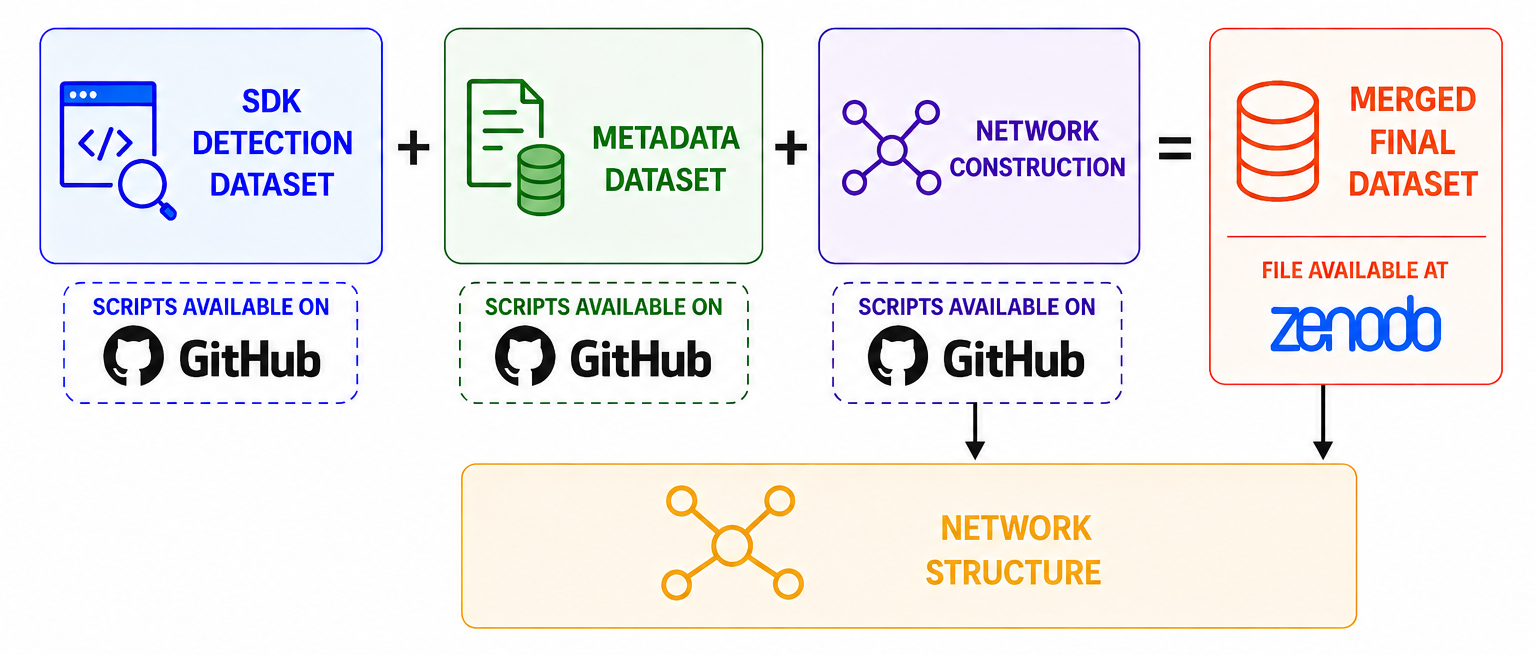}
\caption{{\bf Overview of the released data resources and their relationships.} The SDK extraction outputs, metadata integration pipeline, and network construction procedures are generated through reproducible scripts made publicly available in the GitHub repository. These components are integrated to create the final app-level dataset, which is archived on Zenodo. Starting from the merged dataset, researchers can construct multiple network representations, including the bipartite app-SDK network and one-mode projections such as the app-app and SDK-SDK ones, directly from the publicly available source code.}
\label{fig:data_records_overview}
\end{figure*}

Provider-level identifiers substantially expand the analytical possibilities of the dataset: by aggregating SDKs under common corporate ownership, researchers can move beyond the component level and investigate the structure of the mobile ecosystem at the level of firms; this enables the construction of app-provider, provider-provider, and multi-layer dependency networks, as well as the study of market concentration, technological centralization, and the diffusion of tracking and analytics infrastructures across applications. More broadly, the canonical provider mapping makes it possible to analyze how a relatively small number of corporations extend their presence throughout the Android ecosystem through portfolios of modular software components embedded into third-party apps.\\

Since the dataset preserves APK temporal metadata (subject to the limitations discussed above) together with version identifiers, temporal snapshots of the bipartite ecosystem can be constructed as well. Let $\mathbf{B}^{(t)}$ denote the bipartite matrix associated with a given temporal snapshot indexed by compilation year, release period, or \texttt{dex\_date}: this representation enables longitudinal analyses of SDK adoption, provider diffusion, and technological dependency evolution over time.

\subsection*{Metadata integration}

In addition to SDK detections, the released dataset integrates app-level metadata obtained from the AndroZoo Google Play metadata API. Metadata were retrieved for each sampled package-version pair and stored in both newline-delimited JSON (NDJSON) and flattened tabular formats: the original JSON records were preserved in NDJSON format, while nested fields were flattened into tabular columns using the prefix \texttt{meta\_}; variables prefixed with \texttt{meta\_}, therefore, correspond to flattened Google Play metadata fields. This representation allows the dataset to retain detailed Google Play metadata while remaining directly usable in standard data-analysis environments. Table~\ref{tab:metadata_variables} summarizes the main metadata variables included in the released dataset.

Since Google Play metadata may be collected at a date different from the release date of a specific APK version, the dataset includes the field \texttt{meta\_az\_metadata\_date}, which records the date on which the metadata were acquired from AndroZoo: app-level fields such as download counts, comments, descriptions, developer information, and ratings should, therefore, be interpreted as referring to the metadata acquisition date rather than to the original APK release date; version-specific fields such as version code, upload date, installation size, and APK hashes, instead, describe the corresponding app version when available.

Not all metadata fields are available for every app version: the metadata were linked to the SDK extraction records using deterministic matching procedures based on package names and version-code information whenever available.

\newpage\section*{Data Records}

\subsection*{Directory structure}

The complete resource is distributed across two complementary repositories: a Zenodo archive containing the final processed datasets and a public GitHub repository containing the full reproducible data construction pipeline. The directory structure is organized to separate raw reference files, intermediate outputs, final datasets, validation procedures, and executable source code, thereby facilitating both direct data reuse and end-to-end construction of the dataset.

Figure~\ref{fig:data_records_overview} provides a schematic overview of the released resources. The pipeline begins with two primary inputs: \emph{i)} the SDK detection rule base derived from Exodus Privacy; \emph{ii)} APK-level data and Google Play metadata obtained through the AndroZoo API. These sources are combined to produce the merged app-version dataset, which is subsequently transformed into multiple edge lists and network representations.

The Zenodo archive includes the merged app-version dataset required to reproduce all empirical analyses presented in this study: users can, therefore, download the archive and immediately begin the analysis without requiring direct access to AndroZoo or re-running the extraction pipeline.
The dataset is distributed in CSV format, while selected metadata are additionally provided in NDJSON format: this structure enables researchers either to use the released datasets directly or to reconstruct the complete extraction pipeline on updated APK samples using their own AndroZoo API credentials.

The GitHub repository contains all Python scripts necessary to reconstruct and extend the dataset together with the associated analyses.

\begin{table}[t!]
\centering
\begin{tabular}{lr}
\hline
\textbf{Quantity} & \textbf{Count} \\
\hline
Raw dataset rows & 335,334 \\
Duplicate app-version rows & 615 \\
Unique app-version observations & 334,719 \\
Successfully processed APKs & 334,711 \\
Distinct Android packages & 99,722 \\
Applications containing at least one detected SDK & 71,507 \\
Distinct SDKs empirically observed in the sample & 246 \\
Distinct detectable tracker signatures in the rule base & 277 \\
Raw SDK labels and aliases in provider mapping table & 431 \\
Canonical provider entities & 338 \\
\hline
\end{tabular}
\caption{{\bf Summary of the principal dataset quantities at different levels of aggregation.} The table distinguishes between raw tabular records, unique app-version observations, distinct Android packages, empirically observed SDKs, and the full detectable tracker rule base derived from Exodus Privacy. Let us explicitly notice that the difference between unique app-version observations and successfully processed APKs corresponds to a small number of download or extraction failures. Unless otherwise specified, network representations reported throughout the manuscript are constructed using the subset of apps containing at least one detected SDK.}
\label{tab:dataset_quantity_summary}
\end{table}

\subsection*{GitHub script files}

The source code used to construct the dataset and generate all released network files is publicly available in the GitHub repository \url{https://github.com/auroragorisavellini/android-tracking-sdks-pipeline}. The repository contains the completely reproducible pipeline used for APK sampling, SDK rule extraction, provider mapping, metadata integration, network construction, and technical validation; additional repository files include a \texttt{README.md} file providing step-by-step instructions for reproducing the workflow, a \texttt{requirements.txt} file listing software dependencies, and an open-source software license.
\newpage The principal directory structure is organized as follows:

\begin{verbatim}
android-tracking-sdks-pipeline/
|-- scripts/
|   |-- 01_scrape_exodus_rules.py
|   |-- 02_build_app_sdk_dataset.py
|   |-- 03_build_provider_mapping.py
|   |-- 04_build_final_datasets.py
|   |-- 05_build_network_files.py
|   |-- 06_network_construction_and_analysis.py
|
|-- validation/
|   |-- 01_processing_quality_checks.py
|   |-- 02_provider_mapping_validation.py
|   |-- 03a_sample_representativeness_categories.py
|   |-- 03b_sample_representativeness_game_vs_nongame.py
|   |-- 03c_sample_representativeness_free_vs_paid.py
|   |-- 04_sdk_detection_plausibility.py
|
|-- requirements.txt
|-- README.md
|-- LICENSE
\end{verbatim}

\begin{table*}[t!]
\centering
\begin{tabular}{lrrrrrr}
\hline
\textbf{Bipartite networks} & \textbf{\# Nodes} & \textbf{\# Edges} & \textbf{Edge density} & \textbf{Average degree} & \textbf{Assortativity} & \textbf{Clustering} \\
\hline
app-SDK & 71,753 & 214,421 & 0.0122 & (SDK) 871.63 & -0.387 & -- \\
 & & & & (app) 3.00 & & \\
app-provider & 71,704 & 151,651 & 0.0108 & (provider) 770.82 & -0.357 & -- \\
 & & & & (app) 2.12 & & \\
\hline
\textbf{One-mode projections} & \textbf{\# Nodes} & \textbf{\# Edges} & \textbf{Edge density} & \textbf{Average degree} & \textbf{Assortativity} & \textbf{Clustering} \\
\hline
app-app (sample) & 250 & 23,473 & 0.7542 & 187.78 & -0.082 & 0.924 \\
SDK-SDK & 246 & 4,488 & 0.1489 & 36.49 & -0.419 & 0.796 \\
provider-provider & 197 & 2,551 & 0.1321 & 25.90 & -0.371 & 0.750 \\
\hline
\end{tabular}
\caption{{\bf Summary statistics for the principal network representations derived from the dataset.} The table reports the size and structural properties of the bipartite app-SDK and app-provider networks together with their one-mode projections onto the SDK, provider, and application layers. The app-app projection is constructed from a random sample of 250 applications for illustrative purposes.}
\label{tab:network_summary_statistics}
\end{table*}

The \texttt{scripts/} directory contains the following scripts:

\begin{itemize}
\item \texttt{01\_scrape\_exodus\_rules.py}: downloads and structures SDK detection rules from Exodus Privacy, producing a normalized table of tracker names, namespace-based detection signatures, regular-expression rules, and SDK functional categories;
\item \texttt{02\_build\_app\_sdk\_dataset.py}: draws a reproducible random sample of Android apps from the subset of AndroZoo packages distributed through Google Play, downloads the corresponding APKs using the AndroZoo API, performs static analysis of Dalvik bytecode, extracts application namespaces, detects embedded SDKs using Exodus Privacy rules, and retrieves Google Play metadata associated with each package;
\item \texttt{03\_build\_provider\_mapping.py}: constructs a canonical mapping between detected SDKs and the companies that develop or control them by combining automated web scraping, manual validation, canonicalization procedures, and provider-name normalization;
\item \texttt{04\_build\_final\_datasets.py}: merges SDK detection outputs with flattened Google Play metadata and generates the final app-version dataset together with the app-SDK and app-provider edge lists;
\item \texttt{05\_build\_network\_files.py}: constructs the network files used throughout the study. Starting from the app-SDK and app-provider edge lists, the script generates bipartite app-SDK and app-provider networks together with multiple one-mode projections. In the app-app projection, two apps are connected when they share at least one SDK and edge weights correspond to the number of SDKs they share. In the SDK-SDK projection, two SDKs are connected when they co-occur within the same app and edge weights correspond to the number of apps in which the pair co-occurs. In the provider-provider projection, two companies are connected when SDKs controlled by both providers are embedded into the same app and edge weights correspond to the number of apps in which the two providers co-occur;
\item \texttt{06\_network\_construction\_and\_analysis.py}: constructs bipartite app-SDK and app-provider networks together with multiple one-mode projections, computes descriptive network statistics and structural metrics, and optionally performs community detection using Louvain, Leiden, Stochastic Block Model (SBM), and degree-corrected Stochastic Block Model (dcSBM) approaches\footnote{Community detection based on the SBM and the dcSBM is carried out by running the Python package named DOMINO (\emph{Detection Of Mesoscale structures via INfOrmation criteria})~\cite{Marzi2025DOMINO}.}. The script additionally generates community partitions, summary statistics, serialized graph objects, and publication-ready network visualizations.
\end{itemize}

\begin{figure*}[t!]
\centering
\includegraphics[width=\textwidth]{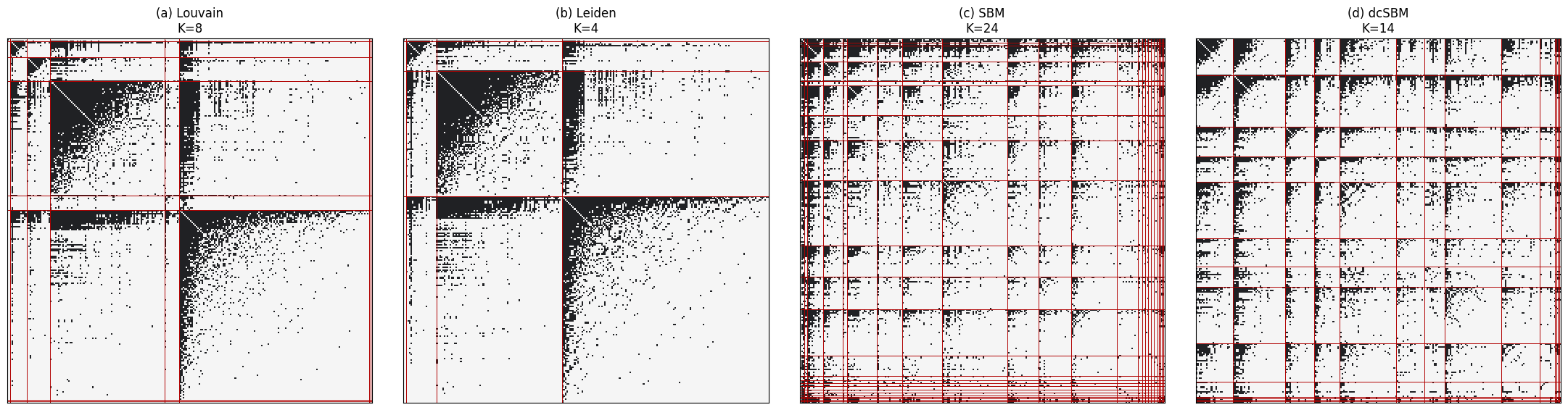}
\includegraphics[width=\textwidth]{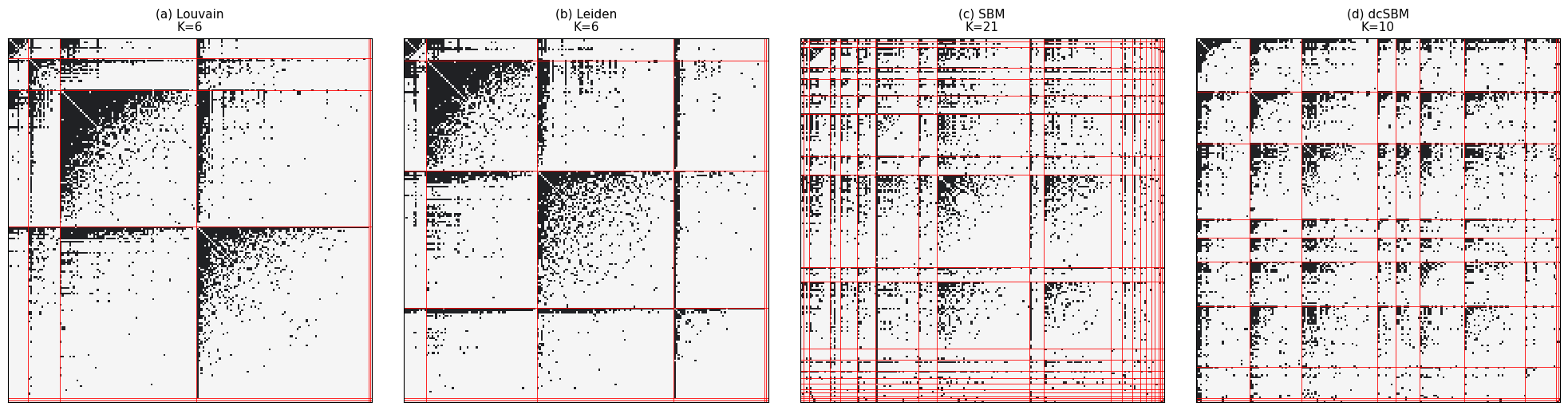}
\caption{{\bf Community structure of the SDK-SDK (top) and provider-provider (bottom) projection network.} Block-ordered adjacency matrices obtained using Louvain modularity maximization, Leiden modularity maximization, the Stochastic Block Model (SBM), and the degree-corrected Stochastic Block Model (dcSBM). Nodes are reordered according to the detected partition and red lines indicate community or block boundaries: the resulting mesoscale organization highlights groups of SDKs that are systematically integrated together across the Android ecosystem as well as groups of providers that frequently appear together within mobile apps.}
\label{fig:sdk_adj_mat}
\end{figure*}

The \texttt{validation/} directory contains scripts used to assess APK processing quality, provider-mapping completeness, sample representativeness, temporal coverage, and SDK detection plausibility through multiple internal and external validation procedures.

All scripts are designed as standalone command-line programs using Python's \texttt{argparse} module, allowing users to reproduce the complete workflow by specifying input and output paths without modifying the source code: together, these files provide a fully documented and version-controlled computational pipeline enabling end-to-end reconstruction of all intermediate and final datasets deposited in Zenodo.

In addition to reproducing the archived data products, the pipeline can be extended and adapted to generate updated APK samples, process future versions of the AndroZoo catalog, or construct additional network representations tailored to specific research questions.

\begin{table}[t!]
\centering
\begin{tabular}{lr}
\hline
\textbf{Metric} & \textbf{Value} \\
\hline
Raw app-version rows & 335,334 \\
Duplicate app-version rows & 615 \\
Unique app-version observations & 334,719 \\
Successfully processed APKs & 334,711 \\
Distinct Android packages & 99,722 \\
APK processing success rate (\%) & 99.998 \\
Metadata merge rate (\%) & 100.00 \\
Applications with at least one detected SDK (\%) & 76.80 \\
Mean detected SDKs per APK & 2.52 \\
Median detected SDKs per APK & 2.00 \\
SDK count inconsistencies & 0 \\
\hline
\end{tabular}
\caption{{\bf Summary statistics for data acquisition and processing quality checks.} The table reports the principal validation metrics used to assess the consistency, completeness, and reliability of the final dataset.}
\label{tab:processing_quality_summary}
\end{table}

\subsection*{Zenodo data files}

The final app-level dataset is publicly archived on Zenodo at \url{https://doi.org/10.5281/zenodo.20313936}. The archive contains the complete release version of the dataset used throughout the study, including SDK detections, Google Play metadata, app identifiers, and auxiliary variables required to reproduce all downstream analyses.

The released dataset is distributed in CSV format and can be directly used to construct all derived edge lists, projected networks, validation outputs, and analytical results using the accompanying open-source pipeline.

The Zenodo archive is organized as follows:

\begin{verbatim}
zenodo_release/
|-- final_app_version_dataset.csv
\end{verbatim}

\subsection*{Final dataset overview}

As described above, the released resource consists of a collection of tabular files documenting third-party SDK adoption in Android applications distributed through Google Play: the core dataset integrates information derived from large-scale static analysis of APK files with application-level metadata obtained through AndroZoo and archived Google Play records. Table~\ref{tab:dataset_quantity_summary} summarizes the principal dataset quantities at different levels of aggregation, including raw rows, unique app-version observations, distinct applications, empirically observed SDKs, and provider entities.

\section*{Data Overview}

Beyond its tabular representation, the dataset can be naturally modeled as a collection of bipartite networks linking Android apps, embedded SDKs, and the companies that develop or control them. The app-SDK bipartite network contains 71,507 apps, 246 distinct third-party SDKs empirically observed in the final APK sample, and 214,421 app-SDK relationships\footnote{The larger edge count of 845,010 reported elsewhere in the manuscript refers to app-version-SDK relationships prior to package-level aggregation: after collapsing APK versions belonging to the same app package, the resulting app-SDK network contains 214,421 unique app-SDK relationships.}; unless otherwise specified, network representations reported throughout the manuscript are constructed at the package level by aggregating APK-version observations belonging to the same package: consequently, the network statistics reported in this section refer to package-level networks rather than to individual APK-version observations. Aggregating SDKs according to corporate ownership yields an app-provider bipartite network comprising 71,507 apps, 197 provider entities, and 151,651 app-provider relationships.

Table~\ref{tab:network_summary_statistics} summarizes the main structural properties of the released network representations, including network size, density, assortativity\footnote{{Assortativity, here, is captured by Newman's coefficient, i.e. the Pearson correlation coefficient between the degrees of connected nodes.}}, and clustering; for bipartite networks, the resulting negative values of assortativity are consistent with a heterogeneous ecosystem characterized by highly prevalent SDKs embedded across large populations of relatively low-degree apps. Besides, in the app-SDK bipartite network, apps embed on average approximately 3.0 SDKs, whereas each SDK appears on average across approximately 871.6 apps: this strong asymmetry reflects the highly concentrated structure of the Android SDK ecosystem, in which a relatively small number of dominant SDKs are embedded throughout a very large population of apps.

Projecting these bipartite structures onto individual node sets produces an SDK-SDK co-occurrence network with 246 nodes and 4,488 edges, and a provider-provider co-occurrence network with 197 nodes and 2,551 edges: in both projections, more than 98\% of nodes belong to a single giant connected component, indicating that the vast majority of SDKs and provider firms belong to a highly interconnected technological ecosystem. For illustrative purposes, we also construct an app-app co-occurrence network based on a random sample of 250 apps, in which two apps are connected whenever they share at least one SDK: this projection exhibits substantial overlap in third-party software dependencies across otherwise unrelated apps and illustrates the dense interconnection structure generated by widely adopted SDK infrastructures.

Figure~\ref{fig:sdk_adj_mat} reports the SDK-SDK and provider-provider projection networks\footnote{For completeness, we also constructed an app-app projection network by connecting apps sharing at least one SDK. Since, however, the resulting projection is extremely dense, its adjacency matrix is largely dominated by the widespread reuse of common SDKs across applications and provides limited insight into mesoscale organization.}, together with the mesoscale partitions identified by Louvain, Leiden, SBM, and degree-corrected SBM. In both projections, nodes are reordered according to the detected partition and displayed together with the corresponding block-ordered adjacency matrices: the resulting structures reveal a non-trivial mesoscale organization characterized by groups of SDKs and providers that systematically co-occur across Android apps. While modularity-based methods (Louvain and Leiden) identify few major communities each of which appears as a core-periphery structure, SBM and dcSBM provide more nuanced block representations.

\begin{figure}[t!]
\centering
\begin{minipage}[c]{0.49\linewidth}
\centering
\includegraphics[height=5.2cm]{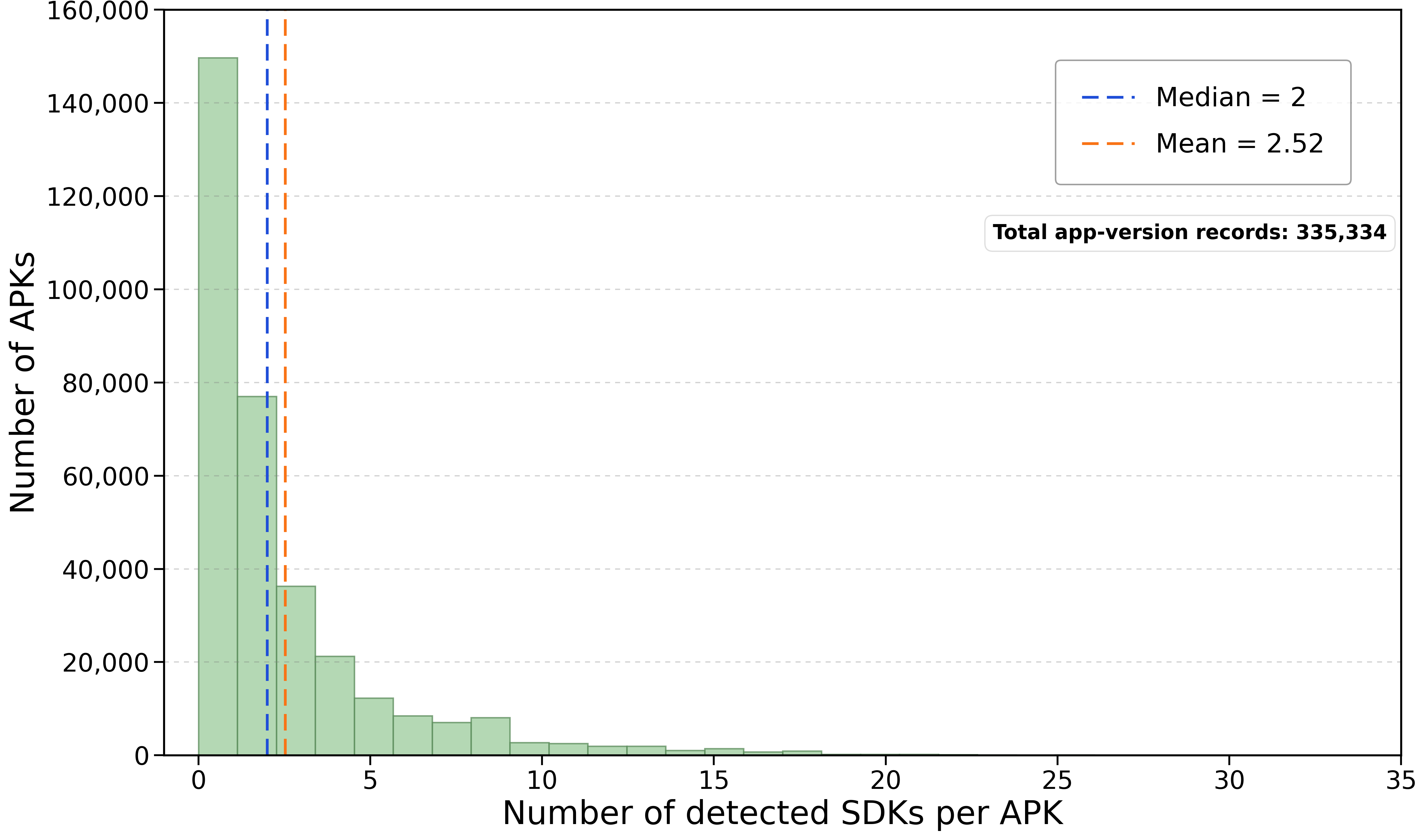}
\end{minipage}
\hfill
\begin{minipage}[c]{0.49\linewidth}
\centering
\includegraphics[height=5.2cm]{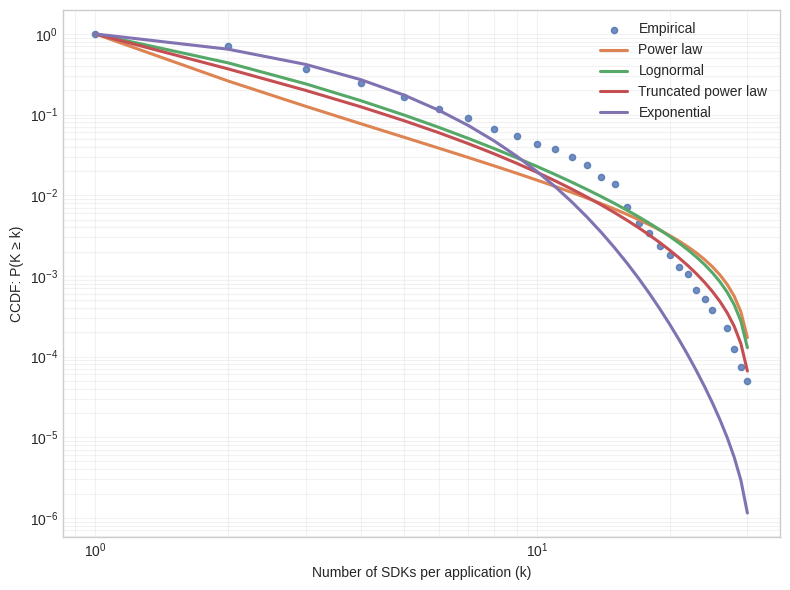}
\end{minipage}
\caption{{\bf Distribution of the number of detected SDKs per APK.} {Left: histogram of the number of tracking SDKs detected within each APK. Most applications embed a relatively small number of third-party tracking SDKs, while a smaller subset contains substantially larger SDK stacks. Vertical dashed lines indicate the sample mean and median. Right: complementary cumulative distribution function (CCDF) of the same distribution shown together with fitted exponential, lognormal, power-law, and truncated power-law models. The figure highlights the strongly right-skewed nature of SDK adoption across apps and the presence of a long upper tail.}}
\label{fig:sdk_count_distribution}
\end{figure}

\section*{Technical Validation}

\subsection*{Data acquisition and processing quality}

We evaluated the reliability of the data acquisition and integration pipeline using multiple internal quality checks. Table~\ref{tab:processing_quality_summary} summarizes the principal validation metrics associated with APK downloading, metadata integration, SDK extraction, and dataset consistency.

The raw dataset initially contained 335,334 app-version rows, including 615 duplicate package-version observations; after deduplication, the dataset contained 334,719 unique app-version observations\footnote{Duplicate package-version rows represent only 0.18\% of the raw merged dataset and do not materially affect downstream analyses because deduplication is performed prior to network construction and statistical analysis.}; among these, 334,711 APKs (99.998\%) were successfully downloaded and processed, with only 8 observations resulting in download or extraction failures. All successfully processed records were matched to the Google Play metadata snapshots archived within AndroZoo, yielding a metadata integration coverage of 100\%.

The static-analysis pipeline detected at least one embedded SDK into 76.8\% of processed APKs: this percentage is computed at the APK-version level. After aggregating APK versions belonging to the same app package, 71,507 distinct packages contained at least one detected SDK and were, therefore, included in the package-level network representations reported throughout the manuscript.

The mean number of detected SDKs per APK was 2.52, with a median of 2, indicating that most applications contain a relatively small number of third-party software dependencies, while a smaller subset embeds substantially larger SDK stacks. Figure~\ref{fig:sdk_count_distribution} illustrates the resulting right-skewed distribution of SDK counts across apps.

Internal consistency checks confirmed perfect agreement between the reported number of detected SDKs (\texttt{sdk\_count}) and the corresponding JSON-encoded SDK lists (\texttt{sdk\_names\_json}), with zero inconsistencies across all processed records.

\subsection*{Sample representativeness by Google Play category}

To assess the external validity of the dataset, we compared the distribution of app categories in our sample with aggregate category statistics reported by 42matters\footnote{\url{https://42matters.com/google-play-statistics-and-trends}}. Using package names from the final dataset, we queried Google Play metadata and successfully retrieved category information for 17,042 applications, corresponding to 23.8\% of all unique apps in the dataset.

This partial coverage is expected for two reasons: first, the sampling strategy based on AndroZoo includes both apps that are currently available on Google Play and apps that were available in the past but have since been removed from the store; second, the mobile app ecosystem is characterized by extremely high turnover, with thousands of applications being introduced and discontinued over relatively short periods of time: as a result, a substantial share of package names present in historical APK repositories can no longer be matched to an active Google Play listing.

Despite this limited match rate, the category composition of the matched applications closely resembles the aggregate distribution reported by 42matters for the Google Play marketplace. Table~\ref{tab:sample_representativeness_categories} reports the comparison for the most prevalent categories:d ifferences between the population shares reported by 42matters and the shares observed in our dataset are generally modest, typically within two-to-three percentage points. For example, Education applications account for 13.4\% of the matched sample compared with 11.26\% in the Google Play population, while Tools represent 6.33\% of the sample versus 7.34\% in the population; similar patterns are observed for Business (5.66\% vs. 8.39\%), Productivity (3.43\% vs. 5.61\%), Food \& Drink (4.27\% vs. 5.22\%), Lifestyle (5.89\% vs. 5.25\%), and Health \& Fitness (5.47\% vs. 4.95\%).

Overall, these results indicate that the dataset captures a broad cross-section of app types and that, conditional on being currently observable on Google Play, the sampled applications are broadly representative of the category composition of the marketplace.

\subsection*{Sample composition: gaming versus non-gaming applications}

As an additional robustness check, we compared the proportion of gaming and non-gaming apps in the dataset with aggregate statistics for the Google Play marketplace reported by 42matters. All Google Play categories whose identifier begins with \texttt{GAME\_} (e.g. \texttt{GAME\_PUZZLE}, \texttt{GAME\_ARCADE}, \texttt{GAME\_SIMULATION}) were collapsed into a single \emph{Game} category, while all remaining categories were grouped into \emph{Non-game}.

Using package names from the final dataset, we successfully matched 17,042 applications to currently available Google Play listings, corresponding to 23.8\% of all unique apps. Because this validation relies on apps that remain observable on Google Play at the time metadata were collected, the matched subset may overrepresent actively maintained or long-lived applications relative to removed or discontinued APKs preserved within AndroZoo.

Among the matched applications, 2,423 (14.2\%) were classified as games and 14,619 (85.8\%) as non-gaming applications: according to the 42matters benchmark, gaming applications represent 11.9\% of the Google Play population, while non-gaming applications account for 88.1\%.

Table~\ref{tab:sample_representativeness_game_nongame} shows that the difference between the sample and population shares is approximately 2.29 percentage points: this relatively small discrepancy suggests that the matched subset broadly reproduces the balance between gaming and non-gaming apps observed in aggregate Google Play statistics.

\begin{table*}[t!]
\centering
\begin{tabular}{lrrrrr}
\hline
\textbf{Category} & \textbf{Population \%} & \textbf{Sample \%} & \textbf{Difference (p.p.)} & \textbf{Population Count} & \textbf{Sample Count} \\
\hline
Education           & 11.26 & 13.44 &  2.18 & 263,870 & 2,290 \\
Business            &  8.39 &  5.66 & -2.72 & 196,621 &   965 \\
Tools               &  7.34 &  6.33 & -1.01 & 172,070 & 1,078 \\
Productivity        &  5.61 &  3.43 & -2.19 & 131,616 &   584 \\
Lifestyle           &  5.25 &  5.89 &  0.64 & 123,195 & 1,004 \\
Food \& Drink       &  5.22 &  4.27 & -0.95 & 122,356 &   728 \\
Health \& Fitness   &  4.95 &  5.47 &  0.52 & 115,941 &   932 \\
\hline
\end{tabular}
\caption{{\bf Comparison between the category distribution of apps in the dataset and aggregate Google Play statistics reported by 42matters.} Population percentages are based on counts reported by 42matters, while sample percentages are computed from the subset of applications for which Google Play category metadata could be retrieved.}
\label{tab:sample_representativeness_categories}
\end{table*}

\subsection*{Sample composition by monetization model}

As a third robustness check, we compared the proportion of free and paid apps in the dataset with aggregate Google Play statistics reported by 42matters. To construct this variable, we used the Google Play metadata field \texttt{meta\_offer.0.micros}, which records the app price in micro-units of the corresponding currency: apps with \texttt{meta\_offer.0.micros = 0} were classified as \emph{Free}, whereas apps with strictly positive values were classified as \emph{Paid}. This field provides a direct and standardized measure of app pricing as reported by Google Play.

The monetization status could be identified for all 71,507 unique apps in the final dataset: among these, 71,506 apps were classified as free and only one app was classified as paid. According to the 42matters benchmark, free apps account for 97.05\% of the Google Play marketplace, while paid apps represent 2.95\%.

Table~\ref{tab:sample_representativeness_free_paid} reports the comparison. The resulting difference reflects a substantial over-representation of free apps and a corresponding under-representation of paid apps within the dataset relative to aggregate Google Play statistics: this pattern is consistent with the economic logic underlying third-party tracking technologies and advertising-supported mobile ecosystems. Free apps frequently rely on advertising, analytics, attribution, and related SDK infrastructures to monetize user attention and behavioral data, making the integration of third-party SDKs a central component of their business model: in this context, user data collection and advertising revenues often constitute the primary economic exchange enabling free access to the app.

By contrast, paid apps generate revenue primarily through direct purchases and subscriptions and, therefore, face comparatively weaker incentives to integrate advertising and behavioral tracking infrastructures: consequently, a dataset specifically designed to capture third-party tracking SDK dependencies is expected to exhibit a strong concentration of free applications.

These results should, therefore, not be interpreted as evidence that the dataset is statistically representative of the entire Google Play marketplace with respect to monetization models; rather, they indicate that the released dataset captures the dominant monetization structure associated with third-party tracking and advertising ecosystems within Android apps.

\begin{table}[t!]
\centering
\begin{tabular}{lrrrrr}
\hline
\textbf{Category} & \textbf{Population \%} & \textbf{Sample \%} & \textbf{Difference (p.p.)} & \textbf{Population Count} & \textbf{Sample Count} \\
\hline
Game      & 11.93 & 14.22 &  2.29 &   279,603 &  2,423 \\
Non-game  & 88.07 & 85.78 & -2.29 & 2,064,795 & 14,619 \\
\hline
\end{tabular}
\caption{{\bf Comparison between the proportion of gaming and non-gaming applications in the dataset and aggregate Google Play statistics reported by 42matters.} All Google Play categories whose identifier begins with \texttt{GAME\_} are grouped into the \emph{Game} category.}
\label{tab:sample_representativeness_game_nongame}
\end{table}

\subsection*{SDK detection plausibility}

To assess the plausibility of the SDK detection pipeline, we compared the prevalence of the most widely used tracking SDKs in our sample with independent statistics reported by Exodus Privacy\footnote{\url{https://reports.exodus-privacy.eu.org/}}. The comparison was conducted using the 21 most prevalent trackers reported by Exodus Privacy: for each SDK, we computed its prevalence in our dataset as the percentage of unique apps (considering only the most recent available APK for each package name) in which the SDK was detected; these estimates were, then, compared with the corresponding prevalence percentages reported by Exodus Privacy.

Table~\ref{tab:sdk_detection_plausibility} reports the comparison. The resulting benchmark indicates broad agreement between the two independently constructed datasets: among the 20 most prevalent SDKs in the Exodus benchmark, 14 also appear among the 20 most prevalent SDKs in our sample, indicating substantial overlap in the dominant technologies identified by the two sources.

The Pearson correlation between prevalence rates is 0.874, revealing a very strong linear association in aggregate adoption patterns: this result suggests that SDKs that are widely adopted according to Exodus Privacy also tend to be widely adopted in our independently collected sample, despite differences in sampling design and application versions. The Spearman rank correlation is 0.566, indicating moderate agreement in the exact ordering of SDKs by prevalence: this result suggests that the strongest agreement is driven primarily by the most prevalent SDKs, whereas differences are more pronounced among medium- and low-prevalence trackers. The difference between Pearson and Spearman correlations suggests that aggregate prevalence levels are strongly aligned across the two datasets, while the precise ordering of intermediate-prevalence SDKs varies more substantially; in practice, a relatively small number of highly prevalent SDKs contribute strongly to the linear association, whereas moderate differences among less prevalent SDKs can lead to noticeable changes in rank ordering. The mean absolute difference in prevalence estimates is 7.66 percentage points.

The comparison shows that the dominant tracking technologies are consistently identified across both datasets: in particular, AppLovin, Facebook Ads, Google AdMob, Google Analytics, Google Firebase Analytics, Google Tag Manager, IAB Open Measurement, ironSource, and OneSignal all appear among the most prevalent SDKs in both sources. Several SDKs additionally exhibit remarkably similar prevalence estimates across the two datasets: for example, AppLovin is observed in 11.73\% of applications in our sample compared with 12\% in Exodus Privacy, while Google Tag Manager appears in 9.91\% and 9\% of applications, respectively; likewise, IAB Open Measurement is detected in 14.17\% of applications in our sample versus 13\% in Exodus Privacy. Table~\ref{tab:sdk_detection_high_agreement} reports selected examples of major SDKs together with their corresponding prevalence estimates in the two independently constructed datasets.

\begin{table}[t!]
\centering
\begin{tabular}{lrrrrr}
\hline
\textbf{Category} & \textbf{Population \%} & \textbf{Sample \%} & \textbf{Difference (p.p.)} & \textbf{Population Count} & \textbf{Sample Count} \\
\hline
Free & 97.05 & 100.00 & 2.95 & 2,274,378 & 71,506 \\
Paid &  2.95 & $<$ 0.01 & 2.95 &    69,250 & 1 \\
\hline
\end{tabular}
\caption{{\bf Comparison between the proportion of free and paid apps in the dataset and aggregate Google Play statistics reported by 42matters.} Apps are classified using the Google Play metadata field \texttt{meta\_offer.0.micros}, where a value of zero indicates a free app and positive values indicate paid apps.}
\label{tab:sample_representativeness_free_paid}
\end{table}

Although Exodus Privacy is a widely used and independent source of tracker statistics, several factors imply that exact agreement between the two datasets is neither expected nor required\footnote{Some discrepancies are, in fact, observed for specific SDKs, such as Amazon Advertisement, Google Crashlytics, and Huawei Mobile Services.}: \emph{i)} the set of applications analyzed by Exodus is not based on a probabilistic sample of the Android ecosystem. Reports are generated on demand by users and members of the community, which may introduce selection biases: popular applications, apps perceived as particularly invasive from a privacy perspective, or applications belonging to specific categories such as games, social media, and utilities may, therefore, be overrepresented; \emph{ii)} the two datasets differ in their temporal and version coverage. Exodus statistics are based on specific versions of apps analyzed at different points in time, whereas our dataset is constructed from a random sample of APKs obtained from AndroZoo and restricted to the most recent available version of each package name: since embedded SDKs can be added, removed, or replaced across software updates, differences in prevalence estimates are naturally expected; \emph{iii)} the two datasets are based on different sampling frames. Our dataset consists of a random sample of APKs available in AndroZoo and subsequently filtered to retain successfully downloaded applications with at least one detected SDK: in contrast, Exodus relies on a dynamic and opportunistically assembled collection of apps submitted for analysis by external users; \emph{iv)} although Exodus employs an established and widely used detection framework, its statistics should not be regarded as an error-free benchmark. As with any large-scale measurement system, the reported results may be affected by false positives or false negatives, delays in updating tracker rules, heterogeneity in the application versions analyzed, and future modifications to the underlying detection procedures.

In this context, the substantial overlap among the most prevalent SDKs and the strong correlation in prevalence rates provide additional evidence that the detection pipeline captures the principal structural features of the Android tracking ecosystem.

\subsection*{SDK-provider mapping validation}

A key contribution of the dataset is the construction of a canonical mapping between third-party Android SDKs and the companies that develop or control them: this mapping is essential for transforming the app-SDK bipartite network into higher-level representations connecting apps to corporate entities, thereby enabling analyses of market concentration, ownership structures, technological intermediation, and inter-organizational dependencies.

The mapping was carried out through a hybrid procedure combining automated web scraping and manual validation: for each SDK listed in the Exodus Privacy tracker rules, we first extracted the tracker website URL and associated metadata from the corresponding Exodus tracker page and then attempted to identify the owning company by parsing website footers, metadata fields, structured JSON-LD information, legal or privacy pages, and textual references associated with the SDK.

Automatically inferred company names were subsequently standardized through a canonicalization procedure designed to harmonize aliases, subsidiaries, spelling variants, legal suffixes, and rebranded entities into unified provider labels (e.g., `Google LLC' and `Google Inc.' mapped to a common \texttt{Google} provider identity); similarity checks based on normalized company names and fuzzy string matching were additionally used to identify candidate duplicates before manual verification.

For SDKs whose ownership could not be reliably inferred automatically, provider assignments were manually validated using publicly available sources including official SDK documentation, developer documentation, corporate websites, privacy policies, acquisition announcements, and industry reports - manual validation was primarily required for SDKs associated with acquisitions, mergers, rebranding events, discontinued products, legacy libraries, or ambiguous product-level naming conventions: provider attribution, therefore, reflects the observable corporate ownership structure at the time the mapping procedure was conducted, i.e. October 2025.

To improve transparency and reproducibility, the provider-mapping pipeline automatically generates an audit table (\texttt{sdk\_provider\_mapping\_final.csv}) containing intermediate extraction outputs, attribution sources, canonicalization fields, manual reconciliation steps, and final provider labels used throughout the matching procedure: all manually validated mappings are, therefore, explicitly documented in the released dataset and publicly available source code (see \texttt{validation/02\_provider\_mapping\_validation.py} and \texttt{code/03\_build\_provider\_mapping.py}), allowing each provider assignment to be independently inspected and reproduced.

To assess the completeness of the resulting mapping, we evaluated coverage at the app, SDK, and edge levels: the provider mapping table contains 431 raw SDK labels and aliases extracted from the Exodus Privacy tracker database, all of which were successfully assigned to a provider, yielding 100\% SDK-level coverage; after canonicalization and normalization, the final detectable SDK rule base contains 277 distinct tracker signatures, all of which were successfully linked to provider entities.

When merged with the app dataset, the resulting app-SDK edge list contains 845,010 app-SDK relationships, all of which were successfully associated with a provider, corresponding to 100\% edge-level coverage; at the app level, all 71,507 apps containing at least one detected SDK were associated with at least one identified provider, also yielding 100\% app-level coverage.

\begin{table}[t!]
\centering
\begin{tabular}{lr}
\hline
\textbf{Metric} & \textbf{Value} \\
\hline
Unique apps in latest-version sample & 71,507 \\
Unique SDKs detected in sample & 246 \\
Exodus SDKs used for benchmark & 21 \\
Top-20 overlap between sample and Exodus benchmark & 14 \\
Spearman rank correlation & 0.566 \\
Pearson prevalence correlation & 0.874 \\
Mean absolute difference (percentage points) & 7.66 \\
\hline
\end{tabular}
\caption{{\bf External plausibility benchmark based on a comparison with tracker prevalence statistics reported by Exodus Privacy.} The analysis uses the latest available version of each app in the sample and compares the prevalence of the most widely detected SDKs across the two independently constructed datasets.}
\label{tab:sdk_detection_plausibility}
\end{table}

Taken together, these results indicate that the SDK-provider mapping is both highly complete and empirically comprehensive. The complete coverage achieved at the app, SDK, and edge levels provides strong evidence that the ownership attribution procedure virtually captures all provider entities embedded into the observed Android tracking ecosystem.

\subsection*{Pipeline robustness}

The dataset is generated through a fully scripted and reproducible pipeline that reconstructs all intermediate and final files from publicly available sources. The workflow is organized into modular Python scripts covering SDK rule extraction from Exodus Privacy, APK sampling and downloading from AndroZoo, static SDK detection, SDK-provider mapping, metadata integration, and final network construction.

All processing stages are implemented programmatically and executed through deterministic procedures using fixed random seeds and immutable SHA256-based APK identifiers: given the same input files and API-access conditions, the pipeline produces identical intermediate and final outputs, thereby ensuring transparent and reproducible dataset construction.

To improve robustness under large-scale execution, the pipeline incorporates resumable checkpointing, incremental writing procedures, and fault-tolerant recovery mechanisms: long-running tasks such as APK downloading, Google Play metadata retrieval, SDK extraction, and provider mapping can, therefore, be interrupted and resumed without recomputing previously completed operations; intermediate outputs are persistently stored at each stage of the workflow, allowing users to inspect, verify, and independently reproduce every transformation step.

Parallel APK downloading and static analysis are implemented through controlled multi-threaded execution compliant with AndroZoo usage recommendations regarding concurrent API requests; retry procedures with exponential backoff are additionally used to mitigate temporary API failures, network interruptions, and incomplete downloads.

The pipeline also performs multiple internal consistency and validation checks throughout the workflow, including duplicate detection, merge completeness verification, SDK count validation, provider coverage assessment, namespace normalization checks, and integrity validation of generated edge lists and network files.

All scripts, dependency specifications, execution instructions, and validation procedures are publicly available in the accompanying GitHub repository, while the generated datasets and derived network files are archived on Zenodo: together, these resources enable the complete data generation process to be independently replicated, inspected, and audited.

\subsection*{Limitations of static SDK detection}

SDK identification in the released dataset is based on static analysis of Android APK bytecode and namespace matching against the Exodus Privacy rule base: static analysis enables scalable and reproducible detection of third-party SDKs across hundreds of thousands of apps without requiring app execution or device instrumentation.

As in previous work on mobile SDK detection~\cite{FealEtAl2020}, static analysis primarily identifies the potential presence of SDK namespaces and code signatures within an app package rather than directly observing runtime behaviour: consequently, some detected SDKs may correspond to inactive, partially integrated, or conditionally executed components; conversely, SDKs relying on aggressive obfuscation, dynamic code loading, runtime dependency injection, or encrypted payload delivery may not always be fully observable through static inspection alone.

\begin{table*}[t!]
\centering
\begin{tabular}{lrrrrr}
\hline
\textbf{SDK} & \textbf{Sample Rank} & \textbf{Exodus Rank} & \textbf{Sample (\%)} & \textbf{Exodus (\%)} & \textbf{Difference (p.p.)} \\
\hline
AppsFlyer                    & 21 & 15 & 2.96  & 8.00  & -5.04 \\
AppLovin (MAX and SparkLabs) & 5  & 10 & 11.73 & 12.00 & -0.27 \\
Google Analytics             & 6  & 11 & 11.25 & 10.00 & +1.25 \\
Google Tag Manager           & 7  & 14 & 9.91  & 9.00  & +0.91 \\
IAB Open Measurement         & 3  & 8  & 14.17 & 13.00 & +1.17 \\
ironSource                   & 13 & 13 & 5.87  & 9.00  & -3.13 \\
OneSignal                    & 12 & 21 & 5.97  & 5.00  & +0.97 \\
Pangle                       & 22 & 18 & 2.87  & 7.00  & -4.13 \\
\hline
\end{tabular}
\caption{{\bf Selected examples of prevalence comparisons between our dataset and Exodus Privacy.} The table reports selected prominent SDKs together with their prevalence estimates in the two independently constructed datasets. Differences are expressed in percentage points (sample minus Exodus).}
\label{tab:sdk_detection_high_agreement}
\end{table*}

The proposed pipeline does not attempt to estimate false-positive or false-negative rates, as no comprehensive ground truth exists for tracker presence in Android applications at this scale\footnote{As an `exploratory' robustness check, a small random subset of APKs was also analyzed using LiteRadar, an independent framework for Android third-party library detection: since LiteRadar detects a substantially broader range of third-party software libraries than Exodus Privacy, the comparison was performed after aggregating detections at the provider level rather than at the individual SDK level. Agreement was quantified using standard provider-level set-overlap measures and remained stable across progressively larger random APK samples (20, 50, and 100 applications), thus providing additional, although qualitative, evidence that both approaches consistently recover the major provider ecosystems despite their different detection objectives. Additional information on LiteRadar is available at \url{https://pypi.org/project/LibRadar/}.}. Instead, the released dataset inherits both the coverage and the limitations of the Exodus Privacy tracker database together with those of static code analysis. Accordingly, the released dataset should be interpreted as a high-confidence approximation of the observable Android tracking ecosystem rather than as an exhaustive census of all tracking SDKs embedded in mobile applications.

Because third-party SDKs constitute one of the principal technological infrastructures through which mobile applications integrate analytics, advertising, attribution, authentication, and user-tracking functionalities, the observed app--SDK relationships provide an informative large-scale proxy for potential data-sharing and tracking infrastructures within the Android ecosystem. Nevertheless, the released dataset should be interpreted as a large-scale representation of observable SDK dependencies embedded into Android apps rather than as a direct measurement of realized runtime behavior.

\subsection*{Temporal aggregation and APK versions}

The released dataset preserves multiple APK versions associated with the same app package rather than collapsing all observations into a single record: as a result, the data support both cross-sectional and longitudinal analyses of SDK adoption and software dependency evolution within the Android ecosystem.

Depending on the research objective, network representations may be, therefore, constructed at different levels of temporal aggregation: in the most disaggregated representation, each app-version pair may be treated as a distinct node, allowing researchers to study the introduction, persistence, and removal of SDK dependencies across successive releases of the same application; alternatively, observations may be aggregated at the package level by retaining only the most recent available APK version or by constructing the union of all SDKs observed across versions associated with the same app. These alternative aggregation strategies may generate substantially different network structures, particularly for highly dynamic applications undergoing frequent SDK integration changes over time.

Since the dataset additionally preserves SHA256 file hashes, package identifiers, version codes, and APK temporal metadata, temporal snapshots of the ecosystem can be also constructed in order to study the evolution of SDK adoption, provider diffusion, and technological dependency across different periods of the Android ecosystem development; however, as discussed above, the \texttt{dex\_date} field provided by AndroZoo may contain placeholder or unreliable timestamps for a subset of APKs and should be, therefore, interpreted as an approximate temporal indicator rather than as an exact release date.

\newpage \subsection*{Error handling, resumability, and reproducibility}

The extraction pipeline was explicitly designed to ensure robustness under large-scale execution: APK downloading and processing were implemented using parallel worker threads, thread-safe writing procedures, and resumable execution checkpoints. Parallelization was intentionally constrained to moderate concurrency levels in accordance with recommendations provided by AndroZoo regarding concurrent API requests: HTTP requests employed retry mechanisms with exponential backoff to mitigate transient network failures and temporary API interruptions.

To support long-running execution in cloud and distributed computational environments, the pipeline incorporated automatic recovery procedures for interrupted storage and execution states: intermediate outputs and processing logs were written incrementally to persistent storage, allowing interrupted runs to resume without recomputing previously processed APKs while additionally facilitating auditability and debugging of large-scale extraction runs.

\begin{table}[t!]
\centering
\begin{tabular}{lr}
\hline
\textbf{Metric} & \textbf{Value} \\
\hline
Raw SDK labels and aliases in mapping table & 431 \\
Mapped SDK labels and aliases & 431 \\
Unmapped SDK labels and aliases & 0 \\
SDK-level mapping coverage (\%) & 100.00 \\
Distinct normalized tracker signatures & 277 \\
Normalized tracker signatures with provider assignment & 277 \\
Normalized SDK coverage (\%) & 100.00 \\
App-SDK edge rows & 845,010 \\
Edges with assigned provider & 845,010 \\
Edge-level provider coverage (\%) & 100.00 \\
Apps containing at least one detected SDK & 71,507 \\
Apps with assigned provider & 71,507 \\
App-level provider coverage (\%) & 100.00 \\
\hline
\end{tabular}
\caption{{\bf Summary statistics for SDK-provider mapping validation.} The table reports the principal technical validation metrics used to assess the completeness and empirical coverage of the canonical mapping between Android SDKs and provider entities.}
\label{tab:sdk_provider_mapping_validation}
\end{table}

Additionally, the pipeline maintains global registries of previously sampled applications to prevent duplicate sampling across independent runs; deterministic random seeds were used throughout the workflow to ensure reproducibility of both the sampling and SDK extraction procedures.

\subsection*{Potential applications of the dataset}

By linking Android apps, embedded SDKs, and their provider companies, the dataset enables analyses of market concentration, data intermediation, innovation diffusion, software supply-chain dependencies, systemic cyber risk, and the structural organization of the mobile application ecosystem; the released app-SDK and provider-level network representations additionally support the study of dependency concentration, ecosystem fragmentation, technological modularity and the diffusion of third-party software infrastructures across large populations of mobile apps.

Since the dataset preserves app metadata, SDK relationships, provider mappings, and temporal information, it can additionally support longitudinal analyses of SDK adoption dynamics, provider expansion, and changes in the organization of mobile software ecosystems over time. The data may also facilitate the development of statistically validated network projections, concentration measures, and ecosystem-level indicators relevant to studies of digital markets and platform governance.

More broadly, the dataset may inform policy-oriented research on market power, digital platform regulation, privacy governance, regulatory compliance, software transparency, and public decision-making concerning competition, security, and and data protection within mobile ecosystems.

\section*{Data Availability}

The app-level dataset used in this study is available at \url{https://doi.org/10.5281/zenodo.20313936}. The archive contains the release version of the dataset necessary to reproduce all downstream analyses presented in this work. Due to redistribution restrictions imposed by AndroZoo, raw APK files and proprietary application binaries are not redistributed as part of the released archive. Researchers wishing to reconstruct the dataset from scratch or extend the analysis to additional APKs must independently obtain access credentials from the AndroZoo platform and provide their own API key.

\section*{Code Availability}

The source code used in this study for APK sampling, SDK detection, provider mapping, network analysis, and technical validation is available at \url{https://github.com/auroragorisavellini/android-tracking-sdks-pipeline}. The repository includes all Python scripts required to reproduce the complete data-processing workflow, together with the derived datasets released as part of this study. In accordance with the AndroZoo redistribution policy, raw APK files and proprietary application binaries are not redistributed as part of the released archive. Public release of the derived datasets and the accompanying reproducibility scripts has been explicitly authorized by the AndroZoo maintainers under these conditions.

\bibliography{bibscientificdata}
\nocite{*}

\section*{Author Contributions}
A.G.S. conceived the study, developed the complete data collection and processing pipeline, performed the data acquisition, APK analysis, SDK detection, provider mapping, dataset integration, technical validation, network analysis, and wrote the manuscript. M.R. contributed to the conceptualization of the study, provided suggestions on validation procedures and robustness analyses, and reviewed the manuscript. T.S. contributed to the network science framework, advised on the construction and analysis of network representations, and reviewed the manuscript. All authors contributed to manuscript revision and approved the final version.

\section*{Competing Interests}

The authors declare no competing interests.

\section*{Acknowledgements}

We are grateful to Marco Alecci for his assistance and responsiveness regarding the AndroZoo platform throughout the data collection process. We also thank the entire AndroZoo community for maintaining and providing this valuable research infrastructure, which enabled the large-scale collection and analysis of Android application data used in this study.

\section*{Funding}

A.G.S. acknowledges financial support from the PhD Program at IMT School for Advanced Studies Lucca.

\section*{Ethics statement}

This study did not involve human participants, human subjects, personal data, animal subjects, or clinical experiments. All data were obtained from publicly available research resources, including AndroZoo and Exodus Privacy, and consist of software artifacts and associated metadata.

\end{document}